 \documentclass[11pt,twocolumn]{article}

 \pdfoutput=1

\usepackage[round]{natbib}   

\usepackage[english]{babel}

\usepackage{graphicx}
\usepackage{amsmath}

\usepackage{enumitem}

\usepackage{float}
\usepackage{ifthen}
\usepackage{hyperref}

\usepackage{titlesec}

\titlespacing*{\section}{0pt}{1.1\baselineskip}{\baselineskip}

\begin{document}

\def\hmath#1{\text{\scalebox{1.5}{$#1$}}}
\def\lmath#1{\text{\scalebox{1.4}{$#1$}}}
\def\mmath#1{\text{\scalebox{1.2}{$#1$}}}
\def\smath#1{\text{\scalebox{.8}{$#1$}}}

\def\hfrac#1#2{\hmath{\frac{#1}{#2}}}
\def\lfrac#1#2{\lmath{\frac{#1}{#2}}}
\def\mfrac#1#2{\mmath{\frac{#1}{#2}}}
\def\sfrac#1#2{\smath{\frac{#1}{#2}}}

\def\pow{^\mmath}



\twocolumn[
\begin{center}
{\bf \Large
\underline{A Third-Law Isentropic Analysis of a Simulated Hurricane.}
}\\
\vspace*{3mm}
{\Large by Pascal Marquet}. \\
\vspace*{3mm}
{\large M\'et\'eo-France CNRM/GMAP / CNRS UMR3589.
 Toulouse. France.}
\\ \vspace*{2mm}
{\large  \it E-mail: pascal.marquet@meteo.fr}
\\ \vspace*{2mm}
{\large  Paper accepted (19th of July 2017) for publication in the}
{\large \it {Journal of the Atmospheric Science}}
\\ \vspace*{2mm}
{\large \url{http://journals.ametsoc.org/doi/abs/10.1175/JAS-D-17-0126.1}}
\vspace*{1mm}
\end{center}

\begin{center}
{\large \bf Abstract}
\end{center}
\vspace*{-3mm}

\hspace*{7mm}
The moist-air entropy can be used to analyze and better understand the general circulation of the atmosphere or convective motions.
Isentropic analyses are commonly based on studies of different equivalent potential temperatures, all of which are assumed to fully represent the entropy of moist air.
It is, however, possible to rely either on statistical physics or the third law of thermodynamics when defining and computing the absolute entropy of moist air and to study the corresponding third-law potential temperature, which is different from the previous ones.
The third law assumes that the entropy for the most stable crystalline state of all substances is zero when approaching absolute zero temperature.

\hspace*{7mm}
This paper shows that the way all these moist-air potential temperatures are defined has a large impact on:
(i) the plotting of the isentropes for a simulation of the Hurricane DUMILE; 
(ii) the changes in moist-air entropy computed for a steam cycle defined for this Hurricane; 
(iii) the analyses of isentropic stream functions computed for this Hurricane; and
(iv) the computations of the heat input, the work function, and the efficiency defined for this steam cycle.

\hspace*{7mm}
The moist-air entropy is a state function and the isentropic analyses must be completely determined by the local moist-air conditions.
The large differences observed between the different formulations of moist-air entropy are interpreted as proof that the isentropic analyses of moist-air atmospheric motions must be achieved by using the third-law potential temperature defined from general thermodynamics.

\vspace*{7mm}

]



 \section{\underline{Introduction}} 
\label{Introduction}
\vspace*{-2mm}

In a recent paper, \citet[hereafter MPZ]{Mrowiec_al_2016} investigated the thermodynamic properties of a three-dimensional hurricane simulation.
The MPZ paper focuses on isentropic analysis based on the conditional averaging of the mass transport with respect to the equivalent potential temperature $\theta_e$.



The methodology employed in MPZ originated in the concept of a thermal or Carnot heat engine applied to hurricanes by 
\citet{Riehl_JAM_1950}, 
\citet{Kleinschmidt_1951},
\citet[hereafter E86, E88, E91]{Emanuel_86,Emanuel_JAS88,Emanuel_91} and
\citet{Gray_JCMM_94} 
and to a study of steam cycle and entropy budgets described in 
\citet[hereafter P11]{Pauluis_2011}. 
In these papers, it is assumed that the equivalent potential temperature $\theta_e$ or the saturation value $\theta_{es}$ are logarithmic measurements of the moist-air entropy, according to 
$s \: \approx \: S_{0m} \: + \: C_{pm} \: \ln(\theta_e)$ or 
$s^{\ast} \: \approx \: S_{0m} \: + \: C_{pm} \: \ln(\theta_{es})$,
which are valid for unsaturated or saturated moist air, respectively.
It is thus assumed that isentropic surfaces are represented by constant values of $\theta_e$ or $\theta_{es}$ despite the terms $S_{0m}$ and $C_{pm}$, which may depend on the water content and may thus vary with space or time, making the link between $s$ and $\theta_e$ or $\theta_{es}$ unclear, at the very least.

The equivalent potential temperatures are sometimes replaced by the moist static energy counterparts 
$C_{pm} \: T + L_v \: q_v + g \: z$ or $C_{pm} \: T + L_v \: q_{sw} + g \: z$, 
where $C_{pm}$ may depend on water vapor or condensed water contents, $L_v$ is the latent heat of vaporization and $q_v$ or $q_{sw}$ is the unsaturated or saturated specific content of water vapor, respectively.

The use of static energies or equivalent potential temperatures as a proxy for the moist-air entropy has been described and discussed at some length in  other studies over the years:
\citet{Malkus_JM_1958}, 
\citet{Miller_JM_1958}, 
\citet{Green_QJ66},
\citet{Darkow_1968},
\citet{Madden_Robi_JAS70,Madden_Robi_JAS72},
\citet{Levine_JAS72},
\citet{Betts_Dugan_73},
\citet{Betts73,Betts_74,Betts_75},
\citet{Emanuel_Rotunno_1987},
\citet{Emanuel_al_87},
E88,
\citet{Emanuel_JAS89},
\citet{Ooyama_JAS90},
\citet{Peixoto_JGR91},
E94,
\citet{Emanuel_JAS95},
\citet{Renno_Ingersoll_96},
\citet{Emanuel_Bister_JAS96},
\citet{Bister_Emanuel_98},
\citet{Pauluis_al_JAS2000},
\citet{Renno_JAS2001},
\citet{Pauluis_Held_2002_I,Pauluis_Held_2002_II},
\citet{Goody_2003},
\citet{Emanuel_2003,Emanuel_2004},
\citet{Bannon_2005},
\citet{Romps2008},
\citet{Smith_QJ08},
\citet{Pauluis_al_2008_Science,Pauluis_al_2010},
\citet{RompsKuang2010},
\citet{Romps2010},
\citet{Emanuel_Rotunno_2011},
\citet{Raymond_JAMES13},
\citet{Pauluis_Mrowiec_2013}.

In the present paper, it is shown that the way in which the moist entropy and the equivalent potential temperatures are defined may lead to opposite results in studies of isentropic processes in hurricanes.
It is shown that, the more the total water varies with space, the more the isentropes differ, leading to large discrepancies in computations and plots of:
1) the isentropes themselves;
2) the changes with space of the equivalent potential temperatures and the moist-air entropies for a thermodynamic heat cycle;
3) the isentropic stream function, $\Psi$, defined in \citet{Pauluis_Mrowiec_2013} and studied in MPZ;
4) the heat input and work function 
studied in 
E88, E91, 
\citet{Emanuel_JAS95},
\citet{Renno_Ingersoll_96},
\citet{Pauluis_Held_2002_I} and 
P11; 
and 
5) the efficiency of a (moist) steam cycle defined in P11.

The definitions and the values of the heat input and work functions, the stream function, the efficiency  and the moist-air entropy itself cannot be subject to arbitrariness.
This paper thus advocates using the absolute value of the moist-air entropy to study the thermodynamic properties of atmospheric processes or to perform isentropic analyses.

It is explained in Appendix~A that the third law of thermodynamics leads to such a physically founded definition of the absolute entropy, which can be computed for all substances.
The third-law values are determined by using experimental calorimetric values of the specific heat and with the hypothesis that the same (zero) value of entropy is reached for the most stable crystalline state of all substances when approaching absolute zero temperature.
Appendix~A also explains that the third-law values are very close to values of entropies determined from theoretical methods based on statistical and quantum physics, for both mono- and poly-atomic molecules.
Accordingly, the term ``third-law'' entropy will hereafter denote either the theoretical or the experimental absolute value of the entropy.

Appendix~B recalls that several applications of the third law of thermodynamics to atmospheric studies already exist: \citet{HaufHoller87}, \citet{Bannon_2005}, and \citet[hereafter M11]{Marquet11}, a process recently improved by \citet{Marquet_Geleyn_2015} and \citet{Marquet15b,Marquet16c,Marquet16b}.
It is thus possible to study the absolute moist-air entropy defined by
$s(\theta_s) \: = \: s_{\rm ref} \: + \: c_{pd} \: \ln(\theta_s)$, where both $s_{ref}$ and $c_{pd}$ are constant and where the third-law potential temperature $\theta_s$ is synonymous with the absolute moist-air entropy.

The paper is organized as follows.
The definitions of most existing equivalent potential temperatures $\theta_e$ or $\theta_{es}$ are recalled in 
section~\ref{subsection_theta}, 
together with the third-law value $\theta_s$.
Several of the existing moist-air entropies are listed in 
section~\ref{subsection_s}, 
including the third-law value $s(\theta_s)$.
A data-set derived from a simulation of the hurricane DUMILE is presented in section~\ref{subsection_Dumile} 
and isentropic surfaces corresponding to $\theta_e$ and $\theta_s$ are computed, plotted and compared in 
sections~\ref{subsection_impact_isentropic_surfaces} 
and \ref{subsection_impact_theta} 
for a cross-section of this Hurricane.
Differences in the associated moist-air entropies are described in section~\ref{subsection_impact_s} 
and it is shown that the two isentropic stream functions computed in section~\ref{subsection_impact_stream_functions} for both $\theta_e$ and $\theta_s$ 
exhibit large differences. 
The heat input and work function are computed in 
section~\ref{subsection_impact_heat_work} 
for a steam cycle in the so-called temperature-entropy diagram.
The efficiencies of such steam cycles are computed and compared in
section~\ref{subsection_impact_efficiency} 
for both $\theta_e$ and $\theta_s$.
Finally, conclusions are presented in section~\ref{section_conclusions}. 




 \section{\underline{Data and method}.} 
\label{section_data_method}
\vspace*{-2mm}

 \subsection{\underline{Moist-air potential temperatures}} 
\label{subsection_theta}
\vspace*{-1mm}

The equivalent potential temperature $\theta_e$ defined by Eq.~(4.5.11) in \citet[hereafter E94]{Emanuel94} can be written as
\vspace*{-4mm}
\begin{align}
 \!\!\!
  {\theta}_{e/E94} 
   & = \: \theta^{\ast} \:
    \exp\! \left( \frac{L_v \: r_v}{c^{\ast}_{pl}  \: T} \right) \:
    {(H_l)}^{- \, \lfrac{R_v \, r_v}{c^{\ast}_{pl}}}
\label{eq_thetae_E94} \: ,
\end{align}
\vspace*{-4mm}

\noindent
where $\theta^{\ast} = T \; {({p_0}/{p_d})}^{R_d/c^{\ast}_{pl}}$,
$T$ is the temperature, $p_0$ is the standard pressure, $p_d$ is the dry-air pressure, and $R_d$ and $R_v$ are the gas constant of water vapor and dry air, respectively.
The specific heat $c^{\ast}_{pl} = c_{pd} + r_t \: c_l$ depends on the values for dry air ($c_{pd}$) and liquid water ($c_l$).
The mixing ratios $r_v$ and $r_t$ represent water vapor and total water, respectively.
$L_v$ is the latent heat of vaporization.
The relative humidity with respect to liquid water $H_l = e/e_{sw}$ is the ratio of the water vapor pressure ($e$) to the saturated value ($e_{sw}$).

The saturated equivalent potential temperatures studied in E86 and E94 will be written as
\vspace*{-1mm}
\begin{align}
  \! \! {\theta}_{es/E86} 
   & = \: \theta^{\ast} \:
    \exp\! \left( \frac{L_v \: r_{sw}}{c^{\ast}_{pl}  \: T} \right)
\label{eq_thetaes} \: , \\
 \!\!\!
  {\theta}_{es/E94}
   & = \: \theta^{\ast}  \:
    \exp\! \left( \frac{L_v \: r_{sw}}{c^{\ast}_{pl}  \: T} \right) \:
    {(H_l)}^{- \, R_v \, r_{sw} / c^{\ast}_{pl}}
\label{eq_thetaes_E94} \: ,
\end{align}
\vspace*{-3mm}

\noindent
where $p$ is the total pressure and $r_{sw}(T,p)$ is the saturation mixing ratio at temperature $T$ and pressure $p$.

The equivalent potential temperature studied in \citet[hereafter B73]{Betts73} can be written as
\vspace*{-1mm}
\begin{align}
  {\theta}_{e/B73} 
   & = \: \theta \; \:
    \exp\! \left( \frac{L_v \: q_v}{c_{pd}  \: T} \right)
\label{eq_thetae_B73} \: ,
\end{align}
where the dry-air potential temperature is $\theta = T \: (p_0/p)^{\kappa}$ with $\kappa = R_d/c_{pd} \approx 0.286$.

The equivalent potential temperature $\theta_e$ on page 1860 of MPZ can be written as
\vspace*{-1mm}
\begin{align}
  \! \! \! \! {\theta}_{e/MPZ} 
   & = \: \theta^{\ast} \:
    \exp\! \left( \frac{L_v \: r_v}{c^{\ast}_{pl}  \: T} \right) \:
    {(H_l)}^{- \, \lfrac{R_v \, r_v}{c^{\ast}_{pl}}}
\label{eq_thetae_MPZ16} \: .
\end{align}
The difference between ${\theta}_{e/MPZ}$ and ${\theta}_{e/E94}$ is that $p_d$ is replaced by $p$ in the Exner function.
The differences between ${\theta}_{e/MPZ}$ and  ${\theta}_{e/E86}$ are that $r_{sw}$ is replaced by $r_v$ and the additional term depending on ${H_l}$ is included.
The differences between ${\theta}_{e/MPZ}$ and  ${\theta}_{e/B73}$ are that $c_{pd}$ is replaced by $c^{\ast}_{pl}$, $q_v$ is replaced by $r_v$ and the additional term depending on ${H_l}$ is included.

Even though the potential temperatures truly considered in MPZ are ${\theta}_{e/E94}$ it is interesting to consider  ${\theta}_{e/MPZ}$  given by Eq.~(\ref{eq_thetae_MPZ16}) in order to demonstrate that the {\it a priori\/} small change of $p_d$ into $p$ may have a large impact on the definition of isentropic processes. 

The third-law based potential temperature defined in M11, \citet{Marquet_Geleyn_2015} and \citet{Marquet15b,Marquet16c,Marquet16b} can be written as
\vspace*{-0mm}
\begin{align}
  {\theta}_{s/M11} 
   & = \: \theta \; \:
    \exp\! \left( - \: \frac{L_v \: q_l + L_s \: q_i}{c_{pd} \: T} \right)
       \:
    \exp\! \left(  \Lambda_r \: q_t  \right)
\nonumber \\
   & 
        \left( \frac{T}{T_r}\right)^{\!\!\lambda \,q_t}
 \!  \! \left( \frac{p}{p_r}\right)^{\!\!-\kappa \,\delta \,q_t}
 \!  \! \left( \frac{r_r}{r_v} \right)^{\!\!\gamma\,q_t}
      \frac{(1\!+\!\eta\,r_v)^{\,\kappa \, (1+\,\delta \,q_t)}}
           {(1\!+\!\eta\,r_r)^{\,\kappa \,\delta \,q_t}}
\nonumber \\
   & 
     \; {(H_l)}^{\, \gamma \, q_l} \;
     \; {(H_i)}^{\, \gamma \, q_i}
     \; {\left( \frac{T_l}{T} \right)}^{\! \lfrac{c_l\, q_l}{c_{pd}}}
     \; {\left( \frac{T_i}{T} \right)}^{\! \lfrac{c_i\, q_i}{c_{pd}}}
\label{eq_thetas} \: ,
\end{align}
where 
$\lambda = c_{pv}/c_{pd}-1 \approx 0.837$,
$\eta = R_v/R_d \approx 1.608$, 
$\varepsilon = R_d/R_v \approx 0.622$,
$\delta = \eta - 1\approx 0.608$
and
$\gamma = R_v/c_{pd} \approx 0.46$.
The term $\Lambda_r = [ \: {(s_v)}_r - {(s_d)}_r \: ]/c_{pd} \approx 5.87$ depends on the third-law values of the reference entropies for water vapor and dry air, and $L_s$ is the latent heat of sublimation.
The water-vapor, liquid, ice and total specific contents $q_v$, $q_l$, $q_i$ and $q_t = q_v + q_l + q_i$ replace the mixing ratios involved in most of the previous formulations.

The water-vapor and dry-air reference entropies ${(s_v)}_r = {s_v}(T_r, e_r)$ and $ {(s_d)}_r =  s_d(T_r, p_r-e_r)$   depend on the reference temperature $T_r=273.15$~K, the total reference pressure $p_r = 1000$~hPa and the water vapor partial pressure $e_r$.
However, it is shown in M11 that ${\theta}_{s/M11}$ defined in Eq.~(\ref{eq_thetas}) is independent of $T_r$ and $p_r$ if the reference mixing ratio $r_r = \varepsilon \: e_r \, / \,  (p_r - e_r) \approx 3.82 $~g~kg${}^{-1}$ corresponds to the saturating pressure value  at $T_r$: $e_r = e_{sw}(T_r) = 6.11$~hPa.

The first two lines of Eq.~(\ref{eq_thetas}) are derived in (40) in M11.
The four terms in the last line of Eq.~(\ref{eq_thetas}) derived in \citet{Marquet16b} are improvements with respect to M11.
They take possible non-equilibrium processes into account, such as under- or super-saturation with respect to liquid water ($H_l \neq 1$) or ice ($H_i \neq 1$) and possible temperatures of rain, $T_l$, or snow, $T_i$, which may differ from the $T$ for dry air and water vapor.

The advantage of the term ${(H_l)}^{\, \gamma \, q_l}$ in Eq.~(\ref{eq_thetas}) compared with ${(H_l)}^{- \, R_v \, r_v / c^{\ast}_{pl}}$ in Eqs.~(\ref{eq_thetae_E94}) or (\ref{eq_thetae_MPZ16}) is that $q_l$ replaces $r_v$ in the exponent, making no impact in clear-air, or under- or super-saturated moist regions (where $H_l \neq 1$ and $r_v$ may be large, but where $q_l=0$) and having less impact in cloud in under- or super-saturated regions (where $H_l \neq 1$ but where, typically, $q_l < r_v$).

The first- and second-order approximations of $\theta_s$ are derived in \citet{Marquet15b,Marquet16c,Marquet16b}, and lead to
\vspace*{-2mm}
\begin{align}
\! \! \! \! 
  {({\theta}_{s})}_1
   & \approx \: \theta \; \:
    \exp\! \left( - \: \frac{L_v \: q_l + L_s \: q_i}{c_{pd} \: T} \right)
       \:
    \exp\! \left(  \Lambda_r \: q_t  \right)
\label{eq_thetas1} \: , \\
\! \! \! \! 
  {({\theta}_{s})}_2
   & \approx \: {({\theta}_{s})}_1 \;
    \exp\! \left[ \: - \: \gamma \: (q_l+q_i) \right]
  \; {\left( \frac{r_v}{r_{\ast}} \right)}^{\! - \: \gamma \: q_t}
\label{eq_thetas2} \: ,\\
\! \! \! \!
  {({\theta}_{s})}_2
   & \approx \: {({\theta}_{s})}_1 \;
    \exp\! \left[ 
     - \: \gamma \;  q_t  \ln{\left( \frac{r_v}{r_{\ast}} \right)}^{\!}
    \: - \: \gamma \: (q_l+q_i) 
    \right]
\label{eq_thetas2_bis} ,
\end{align} 
\vspace*{-3mm}

\noindent
where $r_{\ast} \approx 0.0124$~kg~kg${}^{-1}$.
Both ${({\theta}_{s})}_1$ and ${({\theta}_{s})}_2$ must be multiplied by the last line of Eq.~(\ref{eq_thetas}) if non-equilibrium processes are to be described.

 \subsection{\underline{The moist-air entropies}} 
\label{subsection_s}

The moist-air entropy is computed in M11 from the third law of thermodynamics.
It can be written as
\vspace*{-1mm}
\begin{align}
  s(\theta_{s/M11})
   & \; = \; \:
  s_{\rm ref} \: + \: c_{pd} \: \ln(\theta_s)
\label{eq_s_thetas_M11} \; \; \; \; \mbox{and}
 \\
  \frac{s(\theta_{s/M11})}{q_d}
   & \; = \; \:
  [ \: s_{\rm ref}/q_d \: ] \: + \: [ \: c_{pd}/q_d \: ] \: \ln(\theta_s)
\label{eq_s_thetas_M11_oqd} \: ,
\end{align} 
where both $s_{\rm ref} \approx 1139$~J~K${}^{-1}$~kg${}^{-1}$ and
 $c_{pd} \approx 1004$~J~K${}^{-1}$~kg${}^{-1}$ are constant, making $\theta_s$ a true equivalent of the specific moist-air entropy.
The formulation (\ref{eq_s_thetas_M11_oqd})  is expressed ``per unit of dry air'', in order to be better compared with the entropies computed in other studies such as E94 or MPZ.

Other definitions of ``moist-air entropy'' are derived with either $s_{\rm ref}$ or $c_{pd}$ (often both of them) replaced by quantities that depend on the total-water mixing ratio $r_t$.
This is true in Eq.~(4.5.10) in E94, which can be written as
\vspace*{-1mm}
\begin{align}
  \frac{s(\theta_{e/E94})}{q_d}
   & \; = \; [ \: - \: R_d \ln(p_0)  \: ] \: + \: c^{\ast}_{pl} \: \ln(\theta_{e/E94})
\label{eq_s_thetae_E94} \: ,
\end{align}
where $p_0 \approx 1000$~hPa is a constant standard value.
The division of $s$ by $q_d$ means that the entropy in Eq.~(\ref{eq_s_thetae_E94}) is expressed ``per unit mass of dry air''.
The reference values of entropies disagree with the third law in E94, the consequence being that several terms are missing or are set to zero in Eq.~(\ref{eq_s_thetae_E94}).
These missing terms may impact the specific entropy if $q_t$ varies in space or time, since these missing terms must be multiplied by $q_d = 1 - q_t$ to compute $s$ from $s/q_d$ given by Eq.~(\ref{eq_s_thetae_E94}).
Moreover, since $c^{\ast}_{pl} = c_{pd} + r_t \: c_l$ depends on $r_t$, changes in $s$ and $s/q_d$ cannot be represented by $\theta_{e/E94}$ in Eq.~(\ref{eq_s_thetae_E94}) for varying values of $r_t$.
This prevents $\theta_{e/E94}$ from being a true equivalent of the specific moist-air entropy, $s$, for hurricanes where properties of saturated regions (large values of $r_t$) are to be compared with non-saturated ones (small values of $r_t$).

Similarly, the moist-air entropy defined in section~3 in MPZ can be written as
\vspace*{-1mm}
\begin{align}
 \! \! \! \! \!
  \frac{s(\theta_{e/MPZ})}{q_d}
   & \; = \;  [ \: - \: c^{\ast}_{pl} \: \ln(T_0) \: ] \: + \: c^{\ast}_{pl} \: \ln(\theta_{e/MPZ})
\label{eq_s_thetae_MPZ} \: ,
\end{align}
where $T_0 = 273.15$~K is a constant standard value.
Again, it is an entropy expressed ``per unit mass of dry air'' and the specific heat $c^{\ast}_{pl}$ depends on varying values of $r_t$, twice preventing $\theta_{e/MPZ}$ from being a true equivalent of the specific moist entropy.

To better analyze the impact of $r_t$ on the term $c^{\ast}_{pl}$, and thus on the definition of the moist-air entropy, two kinds of ``saturated equivalent entropy'' are defined.
They are based on the definition of $\theta_{es/E86}$ given by Eq.~(\ref{eq_thetaes}), yielding 
\vspace*{-1mm}
\begin{align}
  \frac{s(\theta_{es/E86})}{q_d}
   & \; = \: c^{\ast}_{pl} \: \ln(\theta_{es/E86})
\label{eq_s_theta_es_E86_cpast}  \; \; \; \; \mbox{and} \\
  \frac{s(\theta_{es/E86})}{q_d}
   & \; = \: c_{pd} \: \ln(\theta_{es/E86})
\label{eq_s_theta_es_E86_cpd} \: .
\end{align}

 \subsection{\underline{The data set for the Hurricane}
 \underline{DUMILE}} 
\label{subsection_Dumile}

\begin{figure}[hbt]
\centerline{\includegraphics[width=0.99\linewidth]{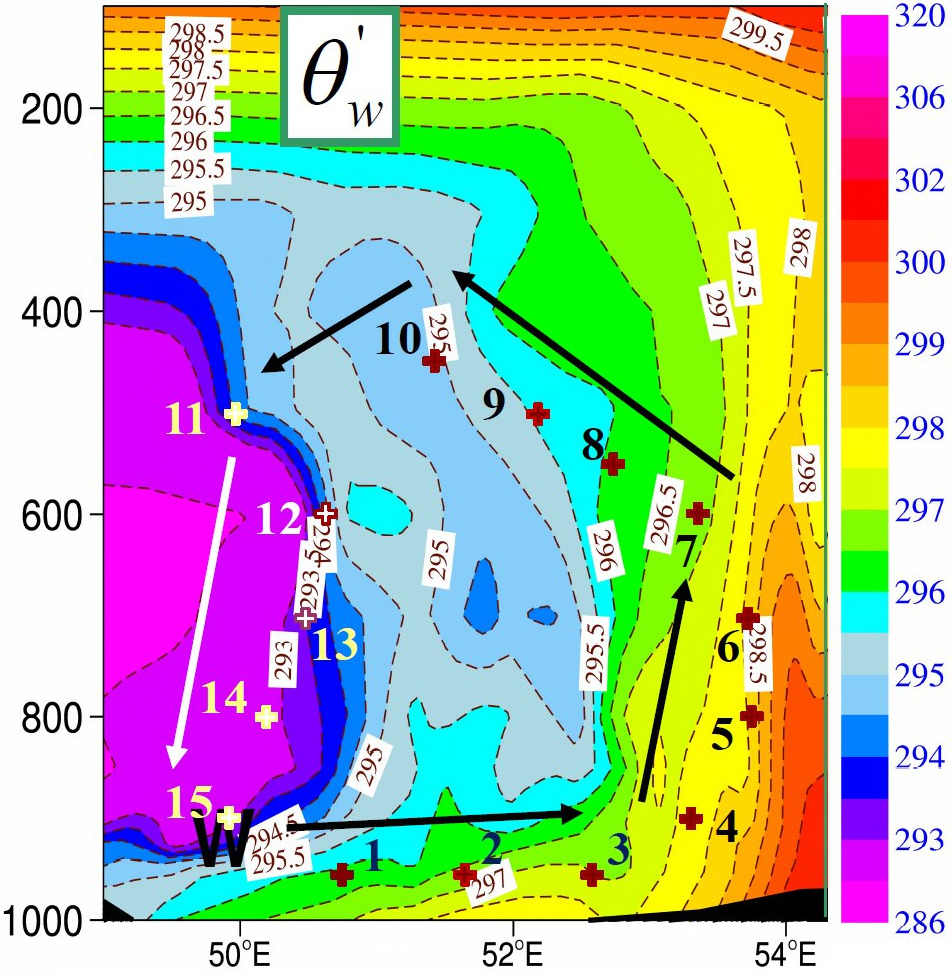}}
\vspace*{-3mm} 
\caption{\it \small A pressure-longitude cross section at latitude $18.5$ south for the Hurricane DUMILE and for the pseudo-adiabatic potential temperature $\theta'_w$ in K.
Data from a $12$~hour forecast for $0000$~UTC 3 January 2013.
The $15$ points (plus signs) represent a steam cycle and were used to build Table~\ref{Table1} and to plot Figs.~\ref{fig6}-\ref{fig7}.
} 
\label{fig1}
\end{figure}

\begin{figure}[hbt]
\centerline{\includegraphics[width=0.99\linewidth]{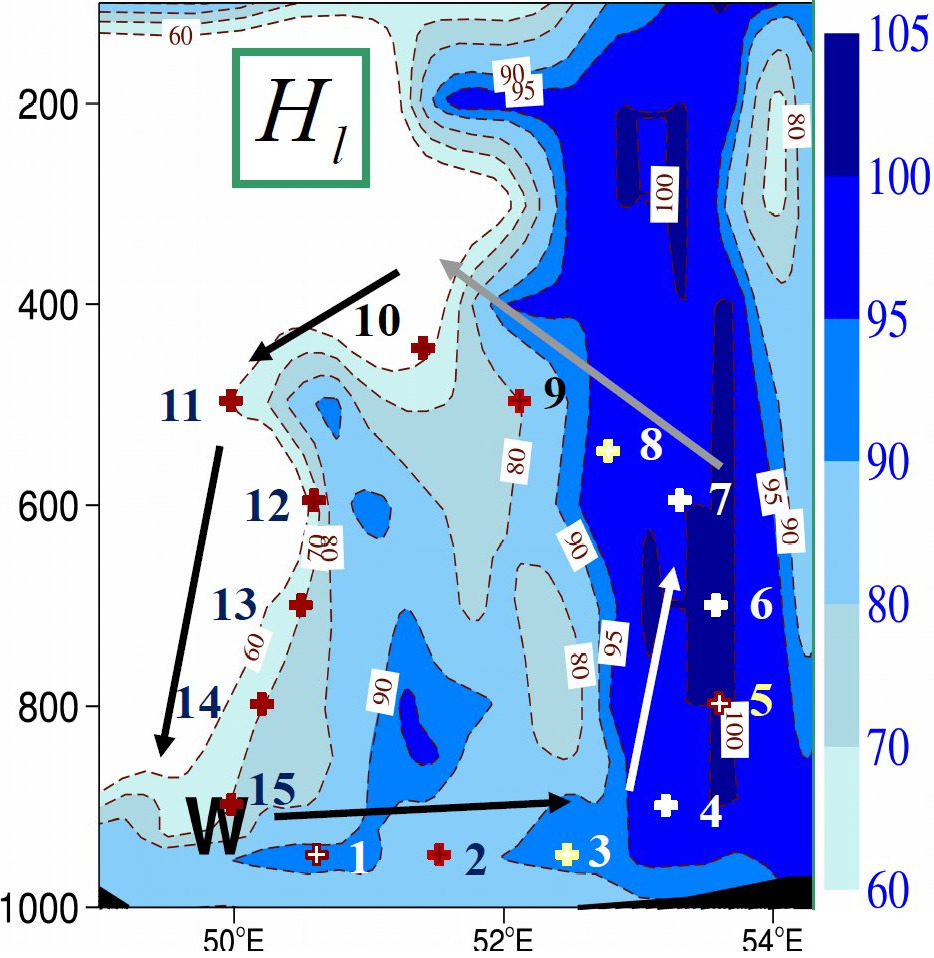}}
\vspace*{-3mm} 
\caption{\it \small As in Fig.~\ref{fig1}, but for relative humidity $H_l$ (\%).
} 
\label{fig2}
\end{figure}

\begin{figure}[hbt]
\centerline{\includegraphics[width=0.99\linewidth]{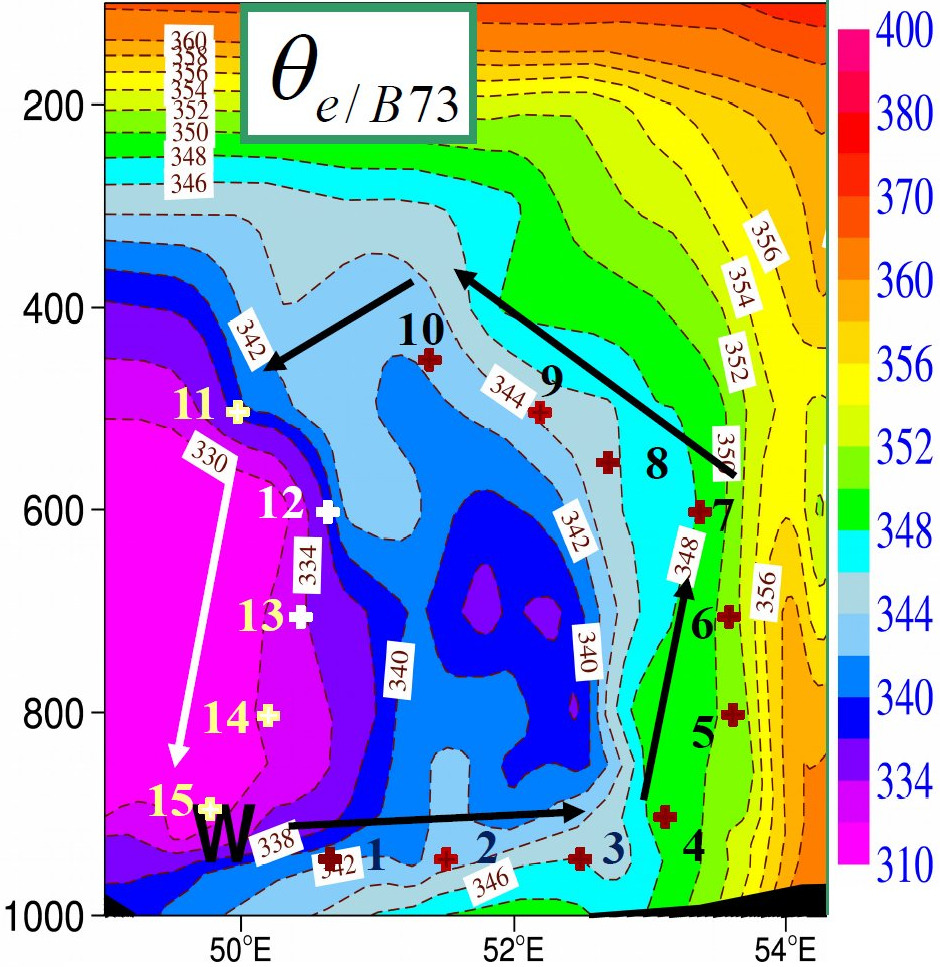}}
\vspace*{-3mm} 
\caption{\it \small As in Fig.~\ref{fig1}, but for the potential temperatures $\theta_{e/B73}$ (K) computed by Eq.~(\ref{eq_thetae_B73}).
} 
\label{fig3}
\end{figure}

The aim of the following sections is to show that varying the values for $r_v$, has significant impact on the definition and the plot of the isentropic surfaces.
To do this, all the moist-air equivalent potential temperatures and the entropies given by Eqs.~(\ref{eq_thetae_E94})-(\ref{eq_s_theta_es_E86_cpd}) are computed for a series of $15$ points selected arbitrarily and plotted in the west-east cross sections indicated on Figs.~\ref{fig1}-\ref{fig4} for the Hurricane DUMILE.

At $0000$~UTC 3 January 2013, the Hurricane DUMILE was located northwest of R\'eunion Island and east of Madagascar, near latitude $18.5$ south and longitude $54.25$ east.
The pressure is used as a vertical coordinate and the black regions close to $1000$~hPa represent the east coast of Madagascar on the left and the center of DUMILE on the right.

The following figures are plotted for a $12$~hour forecast employing the French model ALADIN (Aire Limit\'ee daptation dynamique D\'evelopement InterNational), with a resolution of about $8$~km.
The ALADIN-R\'eunion model is as described in \citet{Montroty_2008} but with two main changes:
(i) the diagnostic turbulence scheme \citep{CBR2000} is based on a $1.5$-order prognostic TKE equation; and
(ii) the shallow convection is based on \citet{Bechtold_2001}.

The ALADIN model is not perfect, owing to the uncertainties concerning the description of the dynamics (non-conservative schemes) and the thermodynamics (more or less accurate parameterizations of microphysics, turbulence and clouds).
Nonetheless, it is assumed that the quality of the ALADIN model is sufficient to describe, in broad terms, relevant spatial variations of the thermodynamic variables.

The two cross sections marked in Figs.~\ref{fig1} and \ref{fig2} are plotted with the pseudo-adiabatic potential temperature $\theta'_w$ and the relative humidity $H_l$.
The use of $\theta'_w$ gives a clear, unambiguous definition of thermal properties, differing from the uncertain and multiple definitions of $\theta_e$ recalled in section~\ref{section_data_method}.\ref{subsection_theta}, which will be compared in later sections.

The $15$ points describe a moist-air steam cycle similar to the one described in P11 and inspired by the Carnot heat engine studied by E86, E88, E91 and \citet{Emanuel_2004}.
The basic thermodynamic conditions ($p$, $T$, $r_v$, $q_l=q_i=0$) of these points are listed in Table~\ref{Table1}.
The dry descent follows a path of almost constant relative humidity between $60$ and $70$~\% (points $11$ to $15$).

The impacts of condensed water and of non-equilibrium terms will not be tested with these $15$ points, chosen such that $T_l=T_i=T$ everywhere and $H_l=H_i=1$ and $q_l=q_i=0$ for the just-saturated regions (points $5$ to $7$).
However, these impacts are expected to be small in comparison with those induced by the large changes in $q_t = q_v$, which vary from $3$ to $17$~g~kg${}^{-1}$.

\begin{table}
\caption{\it \small Pressure (hPa), temperature (K), water-vapor mixing ratios (g~kg${}^{-1}$), relative humidity (relative to liquid water, \%) and pseudo-adiabatic potential temperature (K) for the $15$ points indicated in Figs.~\ref{fig1}-\ref{fig3}.
\label{Table1}
}
\centering
\vspace*{2mm}
\begin{tabular}{cccccc}
\hline
 N   &  $p$  &  $T$     &  $r_v$  &  $H_l$ & $\theta'_w$ \\ 
\hline
$1$  & $950$ & $295.10$ & $16.25$ & $91.9$ & $296.32$  \\ 
$2$  & $950$ & $296.56$ & $16.24$ & $84.1$ & $296.71$  \\ 
$3$  & $950$ & $296.12$ & $17.45$ & $92.6$ & $297.41$  \\ 
$4$  & $900$ & $294.07$ & $17.11$ & $97.5$ & $297.85$  \\ 
$5$  & $800$ & $290.16$ & $15.41$ & $99.9$ & $298.28$  \\ 
$6$  & $700$ & $285.08$ & $12.64$ & $100$  & $298.01$  \\ 
$7$  & $600$ & $278.05$ &  $8.94$ & $98.2$ & $296.95$  \\ 
$8$  & $550$ & $273.57$ &  $6.90$ & $95.9$ & $296.16$  \\ 
$9$  & $500$ & $270.03$ &  $4.87$ & $80$   & $295.58$  \\ 
$10$ & $450$ & $265.38$ &  $2.84$ & $60$   & $294.85$  \\ 
$11$ & $500$ & $268.89$ &  $3.35$ & $60.1$ & $293.90$  \\ 
$12$ & $600$ & $277.15$ &  $5.95$ & $70$   & $294.39$  \\ 
$13$ & $700$ & $282.52$ &  $7.40$ & $70$   & $293.45$  \\ 
$14$ & $800$ & $286.59$ &  $8.49$ & $70$   & $292.18$  \\ 
$15$ & $900$ & $292.28$ & $10.90$ & $70.1$ & $292.85$  \\ 
\hline
\end{tabular}
\vspace*{-4mm} 
\end{table}

The eye-wall and the core of the Hurricane DUMILE are similar to the Figs.~16 and 12 (top) plotted in \citet{Haw_Imb_MWR76} for the Hurricane Inez, where $\theta_e$ was likely computed as an equivalent for $\theta'_w$.
This is a kind of validation of the modeling of the Hurricane DUMILE by ALADIN.
The same high values of $\theta'_w$ observed in the eye of Inez are simulated in Fig.~\ref{fig1} for DUMILE at the lower and upper levels.
The same area of minimum $\theta'_w$ and lower relative humidity is observed in the core region in the $500$ to $700$~hPa layer.
Fig.~\ref{fig3} shows that the general patterns for $\theta_{e/B73}$ are similar to those for $\theta'_w$ shown in Fig.~\ref{fig1}, but with change of labels for the potential temperature units.
This confirms that the isolines defined by constant values of either $\theta'_w$ or $\theta_{e/B73}$ almost overlap.

\section{\bf \underline{Results}.} 
\label{section_results}

 \subsection{\underline{Impact on plotting isentropic surfaces}} 
\label{subsection_impact_isentropic_surfaces}

\begin{figure}[hbt]
\centerline{\includegraphics[width=0.99\linewidth]{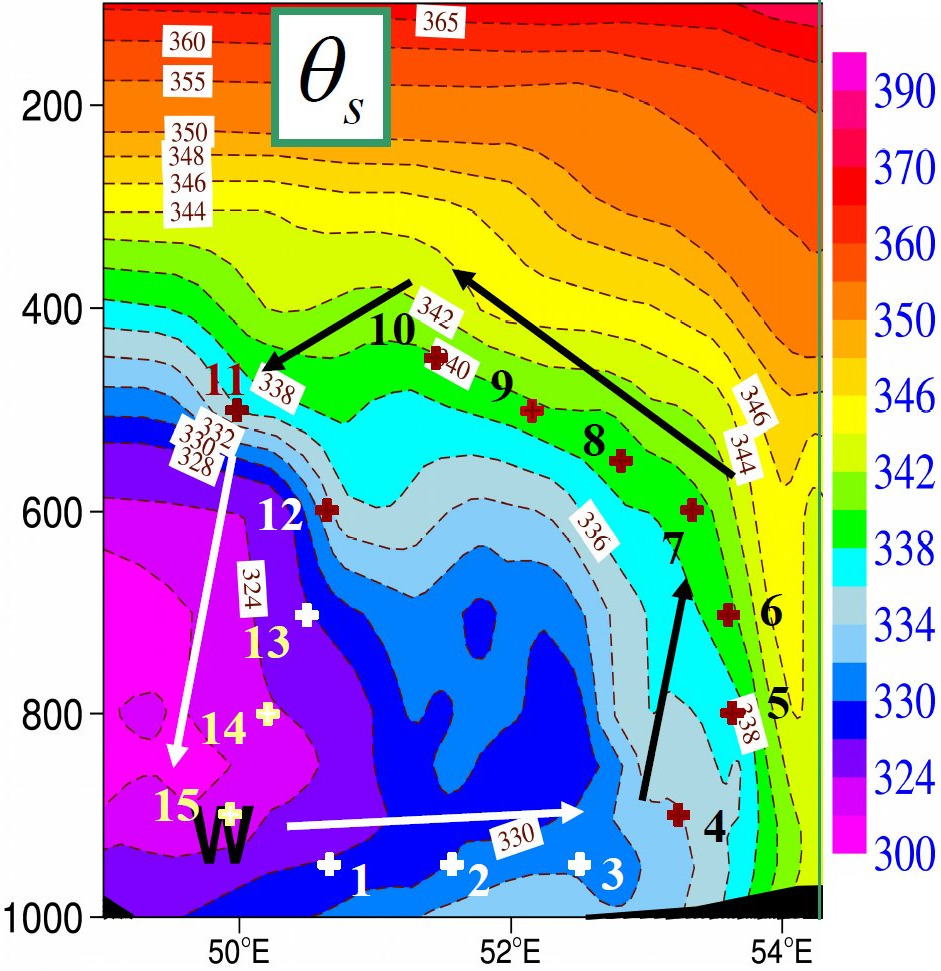}}
\vspace*{-3mm} 
\caption{\it \small As in Fig.~\ref{fig1}, but for the potential temperatures $\theta_{s/M11}$ (K) computed by Eq.~(\ref{eq_thetas}).
} 
\label{fig4}
\end{figure}

\begin{figure*}[hbt]
\centerline{\includegraphics[width=0.9\linewidth]{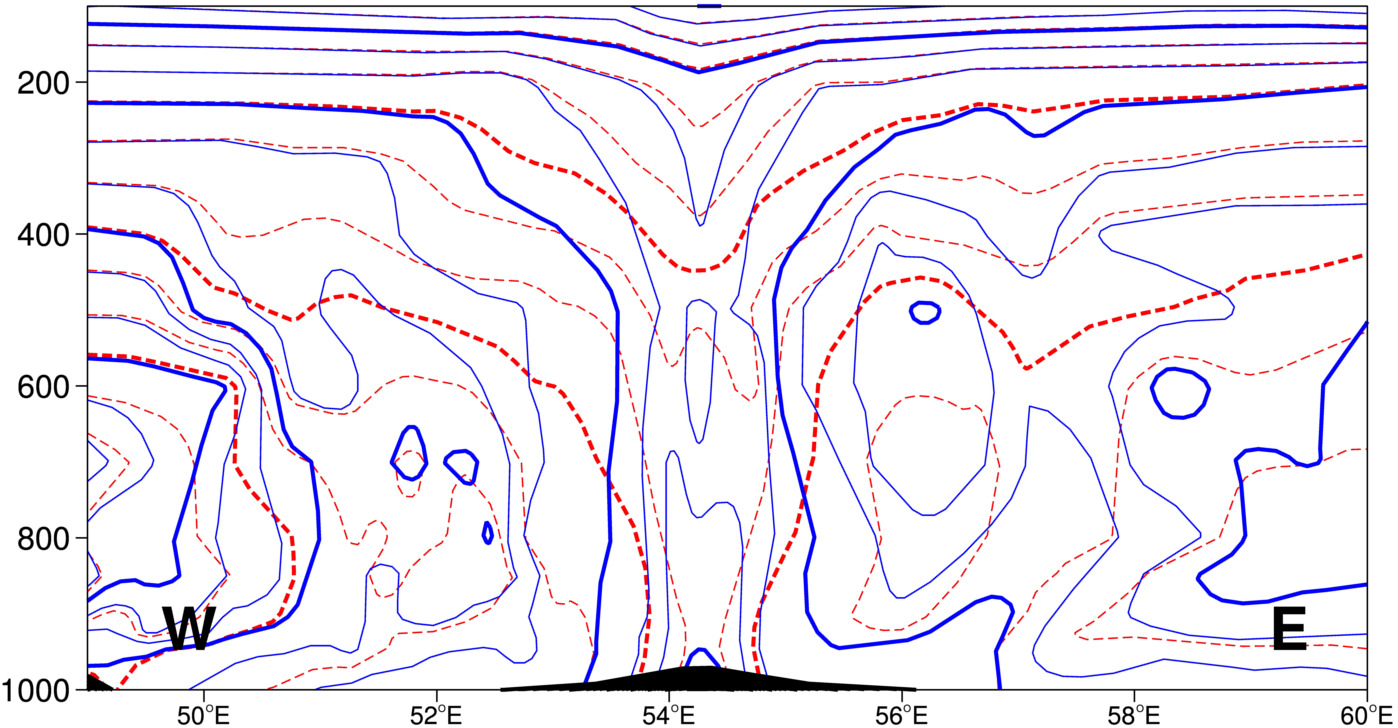}}
\vspace*{-3mm} 
\caption{\it \small As Fig.~\ref{fig1}, but for the complete cross-section and for the potential temperatures $\theta_{e/B73}$ 
(solid blue lines)
and $\theta_{s/M11}$ 
(dashed red lines). 
Isolines are plotted every $2$~K and highlighted every $6$~K.
} 
\label{fig5}
\end{figure*}

The cross section indicated in Fig.~\ref{fig4} for $\theta_s$ exhibits great differences in comparison with Figs.~\ref{fig1} and \ref{fig3} valid for $\theta'_w$ and $\theta_{e/B73}$ and with Fig.~2a for $\theta_{e/MPZ}$ in MPZ.
To facilitate comparisons, the isolines of $\theta_s$ and $\theta_{e/B73}$ are plotted in Fig.~\ref{fig5} for the complete west-east cross-section.

Generally speaking, the isentropes traced with $\theta_{s/M11}$ are smoother than the isolines plotted with $\theta_{e/B73}$.
This can be interpreted as a moderate impact of the total-water content $q_t$ on the moist-air entropy variable
$\theta_{s/M11} \: \approx \: \theta_l \; \exp(6 \: q_t)$,
where the factor 6 is about two-thirds of that included in
$\theta_{e/E94} \: \approx \: \theta_l \; \exp(9 \: q_t)$.
The general aspect of the isolines plotted in Fig.~\ref{fig5} is pretty similar to those that can be observed on polar-equator sections depicted in zonal averages (not shown).

The vertical changes in potential temperatures in the core region are characterized by minimum values of $\theta_s$ close to the surface with increasing values above, up to $100$~hPa.
In contrast, maximum values are observed for $\theta'_w$ and $\theta_{e/B73}$ at $1000$~hPa, with values decreasing up to $600$~hPa and increasing values above.

At some distance from the center, $\theta_s$ exhibits mid-tropospheric minimum values in the layers from $900$ to $750$~hPa while the minimum values of $\theta'_w$, $\theta_{e/B73}$ and $\theta_{e/MPZ}$ are located higher, close to $700$~hPa.
The mixing in $\theta_s$ is greater in the boundary layer below $950$~hPa, where the vertical gradient in $\theta_s$ is smaller than those for $\theta'_w$ and $\theta_{e/B73}$.

The isentropes plotted with $\theta_s$ in Fig.~\ref{fig5} almost coincide with the isolines of $\theta_{e/B73}$ in the dry regions where $q_t$ is small, namely in the high troposphere above the $200$~hPa level and west of the core region (longitudes $<50$~degrees).
The specific humidity, $q_t$, is larger east of the core region because the Hurricane is not symmetric (not shown).
The isentropes are thus different from the isolines of $\theta_{e/B73}$ at all levels below $300$~hPa and for longitudes between $56$ and $60$~degrees.
Consequently, the vertical tilts of the isolines of $\theta_s$ and $\theta_{e/B73}$ can be very different locally, and especially within the eye-walls of the Hurricane (close to or above the points $4$ to $9$, where largest values of $q_t$ are simulated).


The isentropes computed with $\theta_s$ are therefore not compatible with the isolines of $\theta'_w$ or $\theta_{e/B73}$ and the larger the specific content $q_t$ and the relative humidity $H_l$ are in Fig.~\ref{fig2}, the more the isentropes differ in Fig.~\ref{fig5}.
It can be seen in Fig.~\ref{fig5} that the ascending branches of the eye-walls of the Hurricane DUMILE cannot follow both the isentropes plotted with the third-law value $\theta_s$ and the isolines of $\theta_{e/B73}$.
This demonstrates that the different ways of computing the entropy, with either one of the formulations of $\theta_e$ or with $\theta_s$, cannot be true simultaneously.
Only one of them can be applied to plot moist-air isentropes and, thus, to achieve isentropic analyses for the atmosphere.



\begin{figure*}[t]
\centerline{\includegraphics[width=0.8\linewidth]{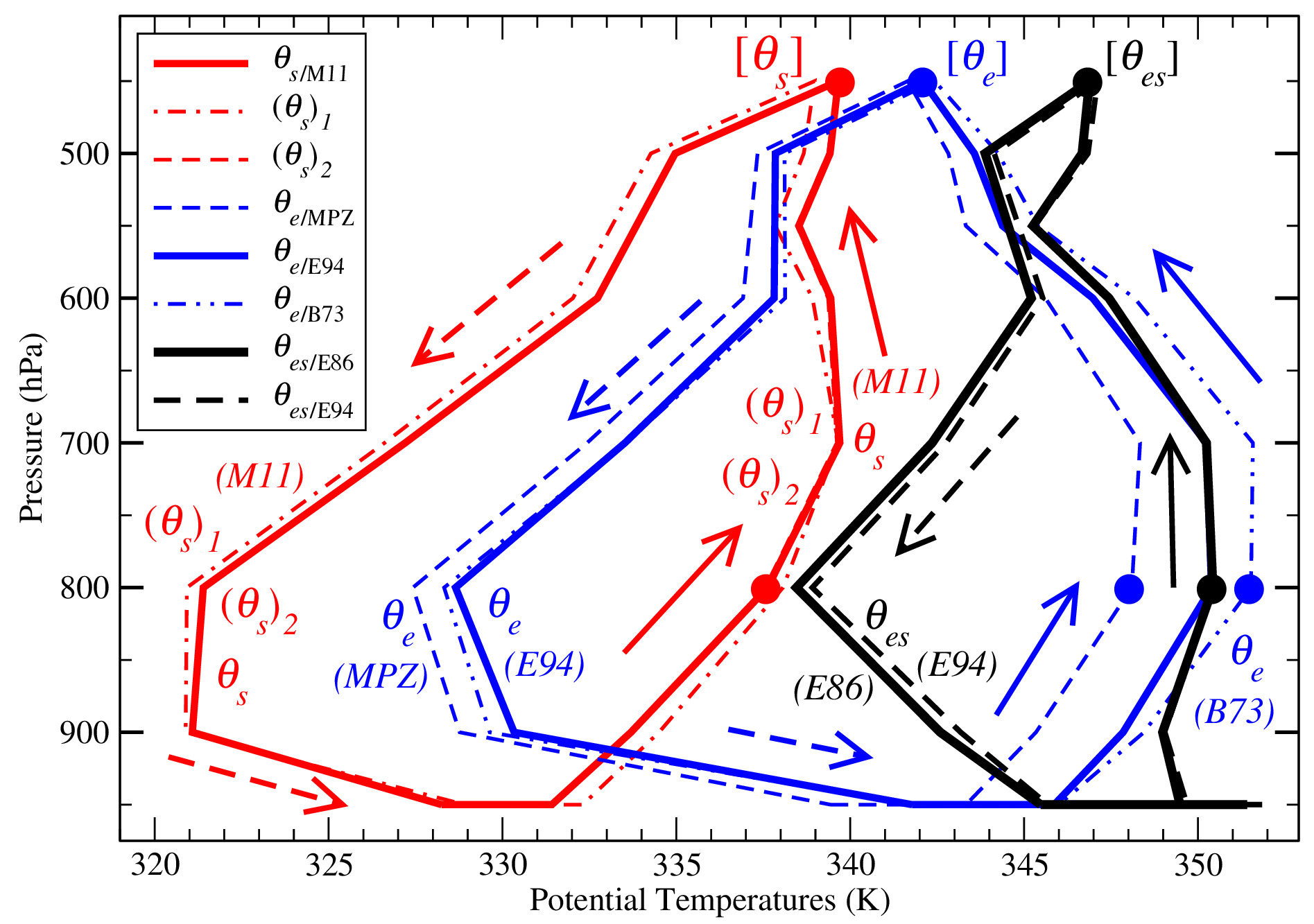}}
\vspace*{-3mm} 
\caption{\it \small The eight potential temperatures given by Eqs.~(\ref{eq_thetae_E94})-(\ref{eq_thetas2})  (K) are plotted against the pressure (hPa) for the $15$ points shown in Figs.~\ref{fig1}-\ref{fig4}.
}
\label{fig6}
\end{figure*}

 \subsection{\underline{Impacts on computing potential}
 \underline{temperatures}} 
\label{subsection_impact_theta}

All moist-air potential temperatures given by Eqs.~(\ref{eq_thetae_E94})-(\ref{eq_thetas2}) are plotted in Fig.~\ref{fig6} and listed in Table~\ref{Table2} (except $\theta_{s1/M11}$).

\begin{table*}
\caption{\it \small
Six potential temperatures (K) computed by Eqs.~(\ref{eq_thetae_E94}) to (\ref{eq_thetas2}).
\label{Table2}
}
\centering
\vspace*{2mm}
\begin{tabular}{ccccccc}
\hline
 N   & $\theta_{s/M11}$ 
     & $\theta_{s2/M11}$ 
     & $\theta_{e/MPZ}$ 
     & $\theta_{e/E94}$ 
     & $\theta_{e/B73}$ 
     & $\theta_{es/E86}$ \\
\hline
$1$  & $328.25$ & $328.27$ & $339.43$ & $341.78$ & $341.74$ & $345.48$ \\ 
$2$  & $329.88$ & $329.87$ & $341.03$ & $343.39$ & $343.12$ & $351.42$ \\ 
$3$  & $331.42$ & $331.42$ & $343.33$ & $345.87$ & $346.01$ & $349.49$ \\ 
$4$  & $333.69$ & $333.67$ & $345.34$ & $347.85$ & $348.45$ & $348.99$ \\ 
$5$  & $337.62$ & $337.55$ & $348.10$ & $350.39$ & $351.53$ & $350.43$ \\ 
$6$  & $339.70$ & $339.62$ & $348.34$ & $350.24$ & $351.58$ & $350.24$ \\ 
$7$  & $339.44$ & $339.37$ & $345.62$ & $346.98$ & $348.18$ & $347.44$ \\ 
$8$  & $338.53$ & $338.47$ & $343.33$ & $344.38$ & $345.40$ & $345.26$ \\ 
$9$  & $339.42$ & $339.37$ & $342.83$ & $343.58$ & $344.56$ & $346.67$ \\ 
$10$ & $339.66$ & $339.63$ & $341.66$ & $342.09$ & $342.43$ & $346.88$ \\ 
$11$ & $334.97$ & $334.94$ & $337.34$ & $337.85$ & $338.11$ & $343.88$ \\ 
$12$ & $332.73$ & $332.69$ & $336.92$ & $337.82$ & $338.13$ & $345.21$ \\ 
$13$ & $327.21$ & $327.18$ & $332.43$ & $333.52$ & $333.55$ & $342.36$ \\ 
$14$ & $321.40$ & $321.40$ & $327.42$ & $328.65$ & $328.32$ & $338.46$ \\ 
$15$ & $321.08$ & $321.09$ & $328.77$ & $330.34$ & $329.63$ & $342.59$ \\ 
\hline
\end{tabular}
\vspace*{-4mm} 
\end{table*}

Three groups of loops are shown in Fig.~\ref{fig6}: 
(i) the three definitions for ${\theta}_{s}$, ${({\theta}_{s})}_1$ and ${({\theta}_{s})}_2$ 
are plotted in red on the left;
(ii) the three equivalent potential temperatures  $\theta_{e/MPZ}$, $\theta_{e/E94}$ and $\theta_{e/B73}$ 
are plotted in blue in the center; and
(iii) the two saturated equivalent potential temperatures $\theta_{es/E86}$ and $\theta_{es/E94}$ 
are plotted in black on the right.

Clearly, ${({\theta}_{s})}_1$, ${({\theta}_{s})}_2$ and ${\theta}_{s}$ remain close to each other with an accuracy of $\pm 0.8$~K for ${({\theta}_{s})}_1$, and with ${({\theta}_{s})}_2$ almost overlapping ${\theta}_{s}$ (errors are smaller than $0.08$~K).
This is new proof that ${({\theta}_{s})}_1$ and ${({\theta}_{s})}_2$ are accurate increasing-order approximations for ${\theta}_{s}$.

The equivalent formulations ${\theta}_{e/MPZ}$, ${\theta}_{e/E94}$ and ${\theta}_{e/B73}$ exhibit larger discrepancies, especially in the warm and moist ascent where differences of $\pm 2.5$~K are observed between $900$ and $700$~hPa.
Moreover, the dry descent for the saturated versions ${\theta}_{es/E86}$ and ${\theta}_{es/E94}$ are $10$~K warmer than that of other definitions of ${\theta}_{e}$, making the loops for ${\theta}_{es/E86}$ and ${\theta}_{es/E94}$ (where $r_v$ is replaced by the larger saturating value $r_{sw}$) much narrower than the others.

The saturated values ${\theta}_{es/E86}$ and ${\theta}_{es/E94}$ remain close to one another.
The differences of about $1$~K in the dry descents can be understood by computing the factor 
$ {(H_l)}^{- \, R_v \, r_{sw} / c^{\ast}_{pl}}$ 
with $H_l \approx 60$~\%, $R_v / c^{\ast}_{pl} \approx R_v/c_{pd} \approx 0.46$
and $r_{sw} \approx 15$~g~kg${}^{-1}$, leading to
$(0.60)^{- 0.46 \: \times \:  0.015} \approx 1.0035$ and to
$340 \times  1.0035 \approx 341.2$~K.

Moderate differences of about $\pm 3.5$~K are observed between the five potential temperatures at high levels, at $450$~hPa where $r_v \approx 3$~g~kg${}^{-1}$ is small. 
The impact is much larger at low level where $r_v > 15$~g~kg${}^{-1}$, with the ascent values of $\theta_s$ about $15$~K colder than  those for $\theta_e$ and $\theta_{es}$ at $950$~hPa.

``Isentropic'' surfaces or regions cannot, therefore, be the same when diagnosed using values of either $\theta_s$, $\theta_e$ or $\theta_{es}$.
The comparisons described in MPZ, which were based on analyses of altitude-$\theta_{e/MPZ}$ diagrams (MPZ's Figs.~4 and 6 to 9), might be invalid since changes in the vertical direction may be of sign opposite the one for $\theta_s$.

The solid disks plotted in Fig.~\ref{fig6} and the corresponding values in Table~\ref{Table1} show that ascents between the levels $800$ and $450$~hPa correspond to a small increase in $\theta_s$ of about $+ 1$~K, while they correspond to a large decrease in $\theta_{e/MPZ}$ of about $- 6$~K.
These differences in signs and in magnitudes between the properties of updrafts and downdrafts are not compatible with the differences in equivalent potential temperatures of about $5$~K described in MPZ (page 1867, right).

Another way to analyze the difference between $\theta_s$ and $\theta_{e/MPZ}$ or $\theta_{e/E94}$ is to consider the gap between descending and ascending values at $800$~hPa:  $\Delta \theta_s \approx 17 $~K and $\Delta \theta_{e/MPZ} \approx \Delta \theta_{e/E94} \approx 21$~K.
The difference is even larger for $\Delta \theta_{e/B73} \approx 24$~K and is much smaller for $\Delta \theta_{es/E86} \approx 12 $~K.

The important consequence of these findings is that changes in moist-air entropy represented by either $\theta_s$ or any of the definitions of $\theta_{e}$ cannot be simultaneously positive or negative, because it would then be impossible to decide whether or not turbulence, convection or radiation processes would increase or decrease the moist-air entropy in the atmosphere.
Moreover, since isentropic processes or changes in entropy must be observable facts, it is impossible to consider that all definitions given by Eqs.~(\ref{eq_thetae_E94})-(\ref{eq_thetas2}) are equivalent; at most only one them can correspond to real atmospheric processes.

\begin{figure*}[ht]
\centerline{\includegraphics[width=0.9\linewidth]{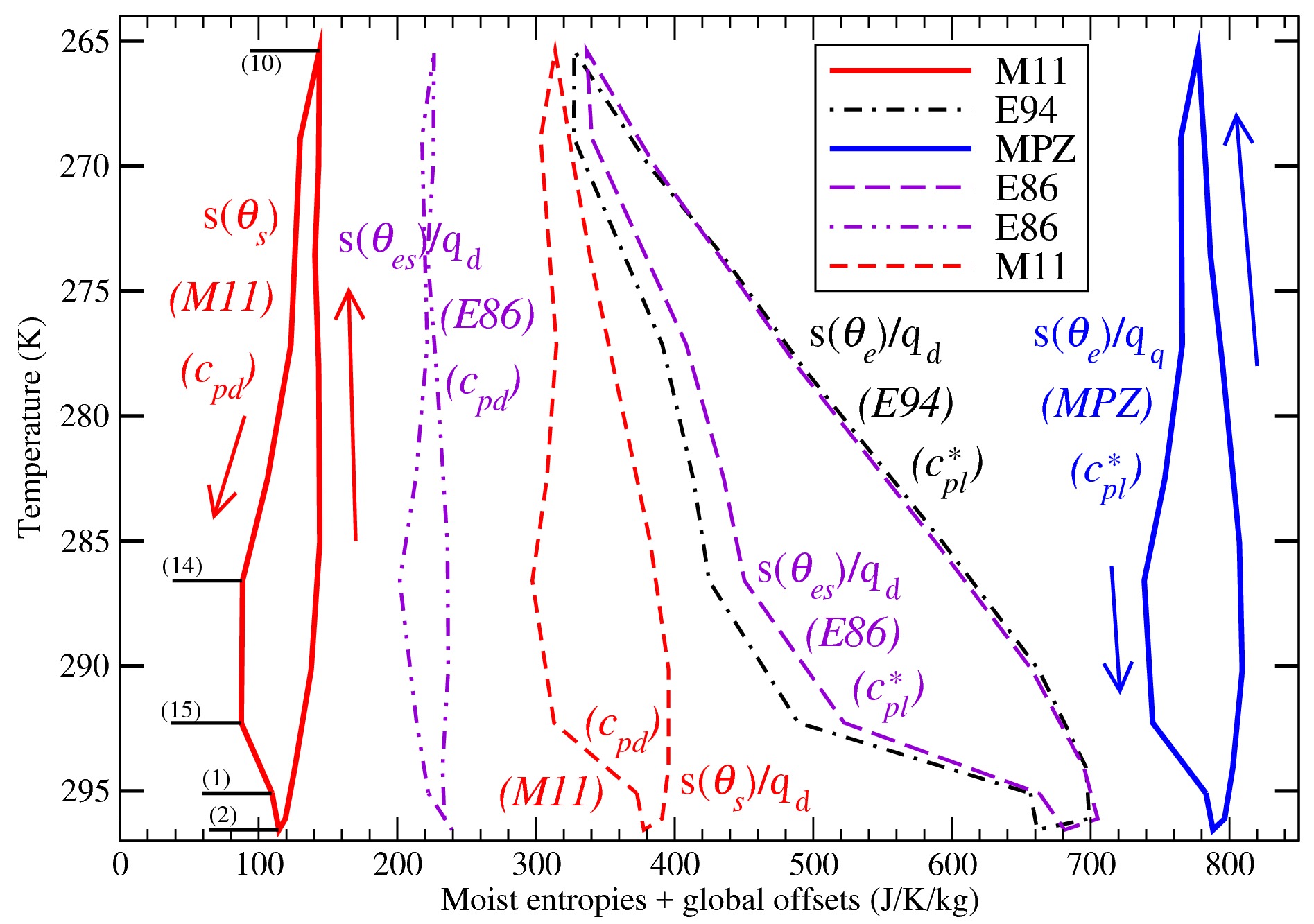}}
\vspace*{-3mm} 
\caption{\it \small The six moist-air entropies (J~K${}^{-1}$~kg${}^{-1}$) defined by Eqs.~(\ref{eq_s_thetas_M11})-(\ref{eq_s_theta_es_E86_cpd}) are plotted in the same temperature-entropy diagram (non-skew Tephigram).
Different global offsets are used for each entropy loop, in order to better separate the loops and facilitate their comparison.
Two arrows and five points are plotted on the first loop for $s(\theta_{s/M11})$ in order to localize the top-level ($10$) and low-level ($14$, $15$, $1$, $2$) values and to describe the anticlockwise rotation of the steam cycle.
The five other loops also rotate anticlockwise with the $15$ points located at the same ordinates.
} 
\label{fig7}
\end{figure*}

\begin{figure}[ht]
\centerline{\includegraphics[width=0.99\linewidth]{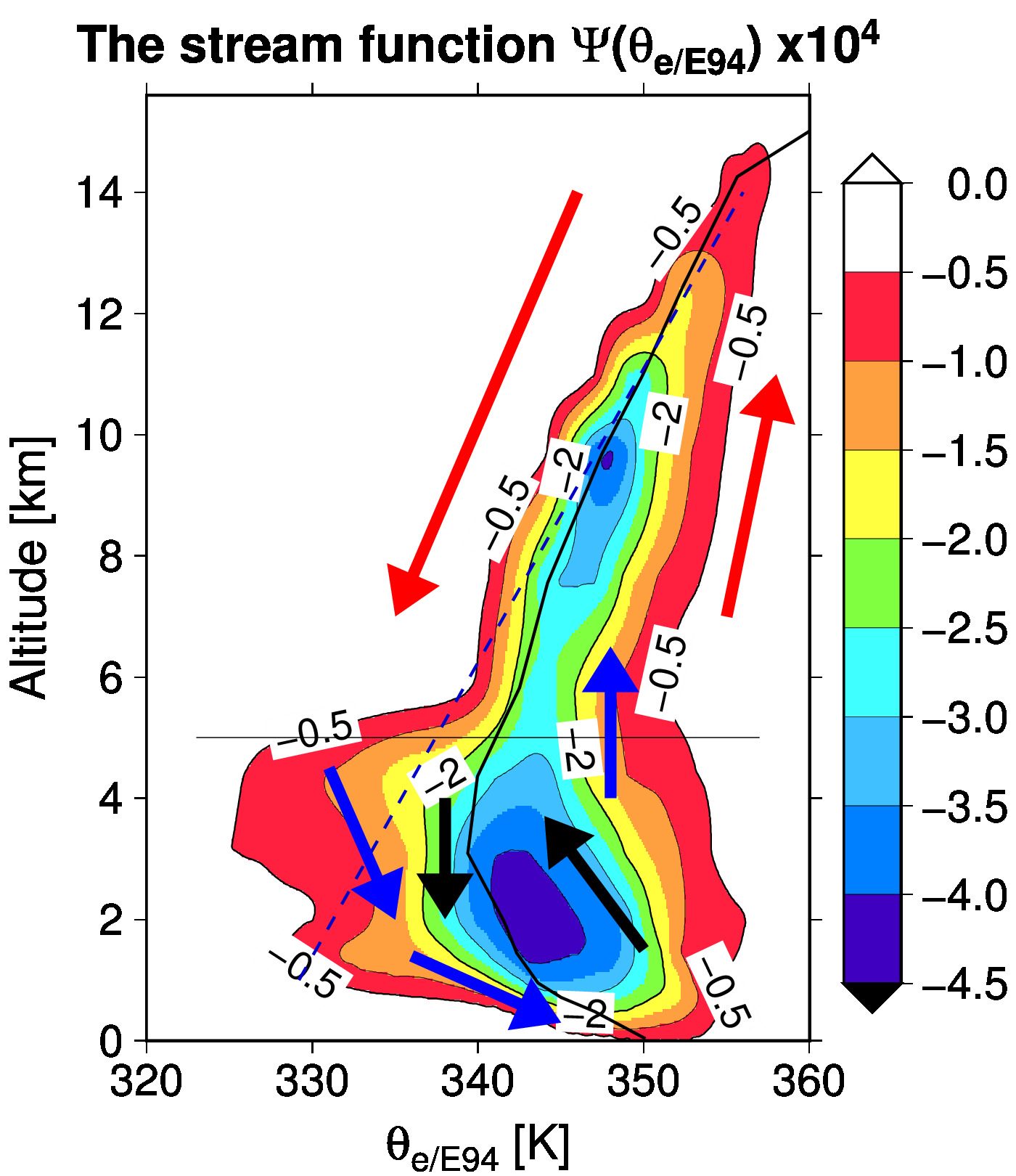}}
\vspace*{-3mm} 
\caption{\it \small The isentropic stream function defined by Eq.~(\ref{eq_Psi_theta}) and by Eq.~(4) in MPZ for ${\theta}_{e/E94}$ ($10^{4}$~kg~m${}^{-2}$~s${}^{-1}$).
The thin solid horizontal line represents the freezing level ($273.15$~K) at about $5$~km.
The solid line shows the vertical profile of the mean values of ${\theta}_{e/E94}$.
} 
\label{fig8}
\end{figure}

\begin{figure}[h]
\centerline{\includegraphics[width=0.99\linewidth]{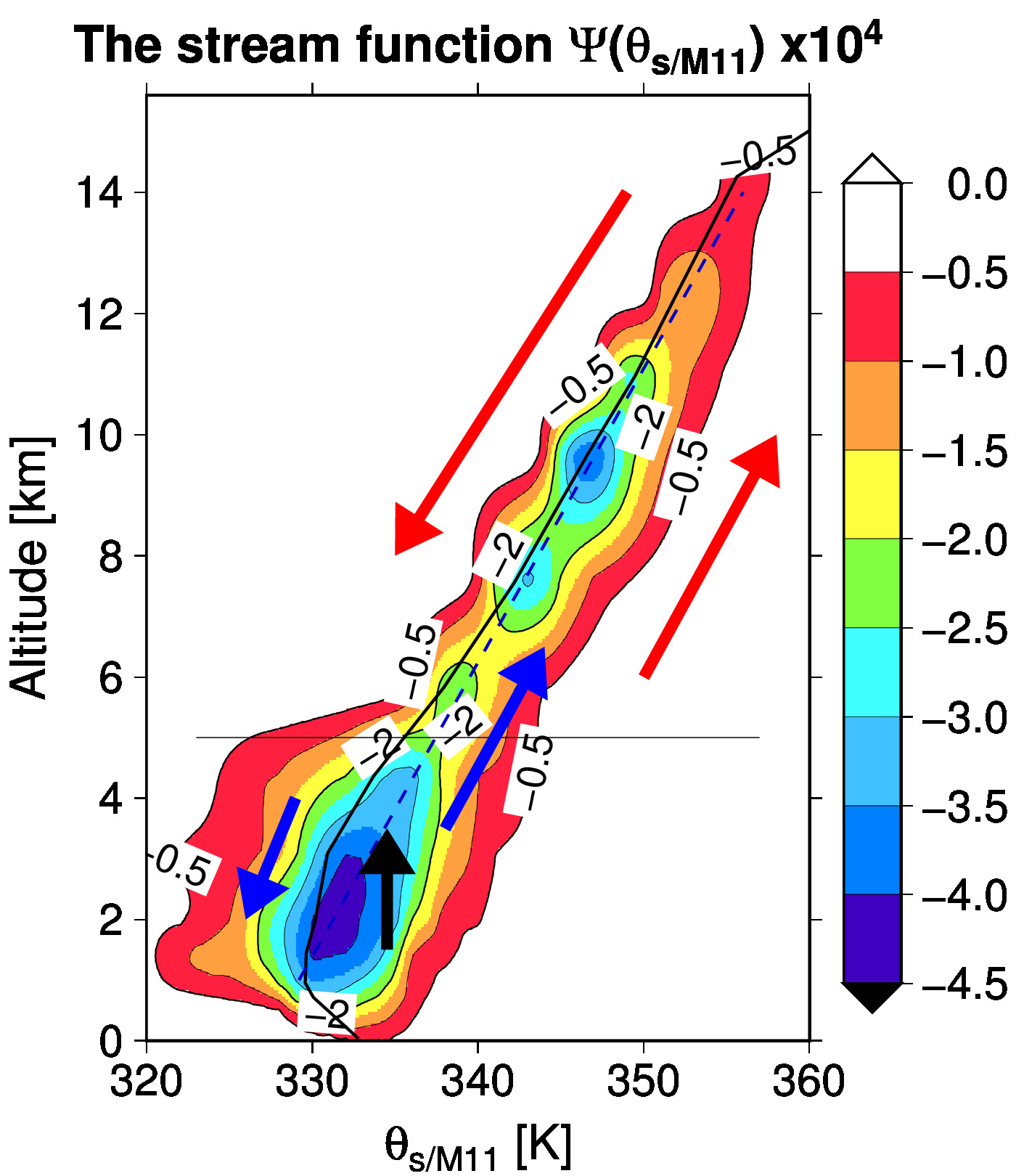}}
\vspace*{-3mm} 
\caption{\it \small As in Fig.~\ref{fig8}, but for ${\theta}_{s/M11}$.
The same dashed straight line connecting the locations of the minima of this stream function have been added in Fig.~\ref{fig8}.
} 
\label{fig9}
\end{figure}

\begin{table*}
\caption{\it \small The moist-air entropies (J~K${}^{-1}$~kg${}^{-1}$) are:  
$s(\theta_{s/M11})$ computed by Eq.~(\ref{eq_s_thetas_M11})
   with an offset of $-6850$~J~K${}^{-1}$~kg${}^{-1}$;
$(s/q_d)(\theta_{es/E86})$ computed by Eq.~(\ref{eq_s_theta_es_E86_cpd}) 
   with an offset of $-5650$~J~K${}^{-1}$~kg${}^{-1}$;
$(s/q_d)(\theta_{e/MPZ})$ computed by Eq.~(\ref{eq_s_thetae_MPZ})
   with an offset of $+550$~J~K${}^{-1}$~kg${}^{-1}$.
The absolute temperature (K) is in the second column in order to follow the loops plotted in Fig.~\ref{fig7} with $T$ as ordinates.
\label{Table3}
}
\centering
\vspace*{2mm}
\begin{tabular}{ccccc}
\hline
 N   &  $T$	& $s(\theta_{s/M11})$  & $(s/q_d)(\theta_{es/E86})$ & $(s/q_d)(\theta_{e/MPZ})$ \\ 
\hline
$1$  & $295.10$ & $109.6$ & $222.4$ & $783.1$  \\ 
$2$  & $296.56$ & $114.6$ & $239.5$ & $788.1$  \\ 
$3$  & $296.12$ & $119.3$ & $234.0$ & $796.5$  \\ 
$4$  & $294.07$ & $126.1$ & $232.6$ & $802.5$  \\ 
$5$  & $290.16$ & $137.9$ & $236.7$ & $809.3$  \\ 
$6$  & $285.08$ & $144.1$ & $236.2$ & $807.2$  \\ 
$7$  & $278.05$ & $143.3$ & $228.1$ & $795.2$  \\ 
$8$  & $273.57$ & $140.6$ & $221.8$ & $786.4$  \\ 
$9$  & $270.03$ & $143.2$ & $225.9$ & $782.9$  \\ 
$10$ & $265.38$ & $143.9$ & $226.5$ & $777.5$  \\ 
$11$ & $268.89$ & $130.0$ & $217.7$ & $765.0$  \\ 
$12$ & $277.15$ & $123.2$ & $221.6$ & $766.0$  \\ 
$13$ & $282.52$ & $106.4$ & $213.3$ & $753.4$  \\ 
$14$ & $286.59$ &  $88.4$ & $201.8$ & $738.5$  \\ 
$15$ & $292.28$ &  $87.4$ & $214.0$ & $744.7$  \\ 
\hline
\end{tabular}
\vspace*{-4mm} 
\end{table*}

 \subsection{\underline{Impacts on computing moist-air} 
 \underline{entropies}} 
\label{subsection_impact_s}

The issue of computing the relevant moist-air entropy is even harder than choosing one of the equivalent potential temperatures studied in the previous section, namely either $\theta_s$ or one of the versions of $\theta_e$ or $\theta_{es}$.
Since the aim of E86, E88, E91, \citet{Emanuel_2004}, \citet{Pauluis_al_2010}, P11 or MPZ is to analyze meteorological properties in moist-air isentropic coordinates, a comparison of values of the moist-air entropy itself is needed.
Let us therefore plot, in the temperature-entropy diagram of Fig.~\ref{fig7}, the values of the six moist-air entropies considered in section~\ref{section_data_method}.\ref{subsection_s}, with three of them listed in Table~\ref{Table3}.

The differences between the loops (Carnot or steam cycles) are large.
Some of the loops are very narrow, whereas others are wide and have a flared shape (a large gap between ascending and descending regions).
Some of the loops are almost vertical (with a small change of less than $\pm 20$~J~K${}^{-1}$~kg${}^{-1}$ in entropy between the surface and the upper air), whereas others exhibit a pronounced tilt (with a large decrease of $400$~J~K${}^{-1}$~kg${}^{-1}$ between the warm, moist regions and the cold, dry ones).
The third-law entropy $s(\theta_s)$ increases between points 2 and 10, whereas all the other formulations lead to a decrease between these two points.

Therefore, since the moist-air entropy is a state function, and since isentropic processes or changes in entropy must be observable facts, it is impossible to consider that all definitions given by Eqs.~(\ref{eq_s_thetas_M11}) to (\ref{eq_s_theta_es_E86_cpd}) are equivalent: at most only one of them can correspond to real atmospheric processes.
Otherwise, it would be possible to modified the second law by imposing arbitrarily an increase or a decrease in entropy for an ascending parcel of moist-air.

 \subsection{\underline{Impact on isentropic stream-functions}} 
\label{subsection_impact_stream_functions}

An isentropic stream function $\Psi(\theta_{e/E94} ; z)$ is studied in \citet{Pauluis_Mrowiec_2013} and MPZ and the alternative third-law value $\Psi(\theta_{s/M11} ; z)$ is tested in this section.
The isentropic stream function defined in MPZ can be written as
\vspace*{-1mm}
\begin{align}
 \!\! \Psi(\theta_{e/s} ; z)
   & = \: \int^{\theta_{e/s}}_{\theta_{min}}
          \int^{+\infty}_0 \!
           < \rho \: w >\!(r,\theta,z) \: \; dr \; d\theta
\label{eq_Psi_theta} \:  ,
\end{align}
where $\theta_{e/s}$ stands either for $\theta_{e/E94}$ or $\theta_{s/M11}$, respectively.
The isentropic mass flux $< \rho \: w >\!(r, \theta, z)$ is computed according to
\vspace*{-1mm}
\begin{align}
\! \! \!
< \rho \: w >
   & =  \iint_A
          \rho \: 
          \left[\: w - \overline{w}(z) \:\right] \:
          D(\theta) \:
          D(r)
          \: \; dx \; dy
\label{eq_correl_rho_w} \: .
\end{align}
The first rectangular function $D(\theta)$ is different from zero and equal to 
$1/ \Delta \theta = 1$K~${}^{-1}$
for $| \theta - \theta'(x,y,z) | < \Delta \theta / 2$
and for $\theta = \theta_{e/E94}$ or $\theta_{s/M11}$ varying from 
$\theta_{min} = 320$ to $\theta_{max} = 360$~K in steps of $\Delta \theta$ ($41$ bin values).
The second rectangular function $D(r)$ is different from zero and equal to 
$1/ \Delta r = 1/15$~km~${}^{-1}$
for $| r - r'(x,y,z) | < \Delta r / 2$
and for $r$ varying from $7.5$ to $487.5$~km in steps of $\Delta r$ ($33$ bin values).
The radial distances $r$ and $r'(x,y,z)$ are computed with respect to the center of the Hurricane DUMILE and the average value $\overline{w}(z)$ is computed for each vertical level for $r<495$~km.




The diagram plotted in Fig.~\ref{fig8} for $\Psi(\theta_{e/E94} ; z)$ and for the Hurricane DUMILE is similar to the one plotted in Fig.~1(b) in \citet{Pauluis_Mrowiec_2013} and in Fig.~7(a) in MPZ for an idealized hurricane.
The locations of the minimum values of the streamlines correspond to zero mean vertical mass flux and they are roughly aligned with the mean values of $\overline{\theta}_{e/E94}(z)$ (solid line).

The arrows give some hints on the average anti-clockwise circulations.
Values of $\theta_{e/E94}$ are conserved only for the descending branch between $4$ and $2$~km and for values close to $338$~K.
Values of $\theta_{e/E94}$ decrease in the ascending branch below the freezing level at $5$~km and they increase above this level up to the top of the troposphere.

The diagram plotted in Fig.~\ref{fig9} for the third-law stream function $\Psi(\theta_{s/M11} ; z)$ exhibits large differences with Fig.~\ref{fig8}, in particular below the freezing level and for the larger values of $q_t$.

\begin{figure*}[t]
\centerline{
\includegraphics[width=0.4\linewidth]{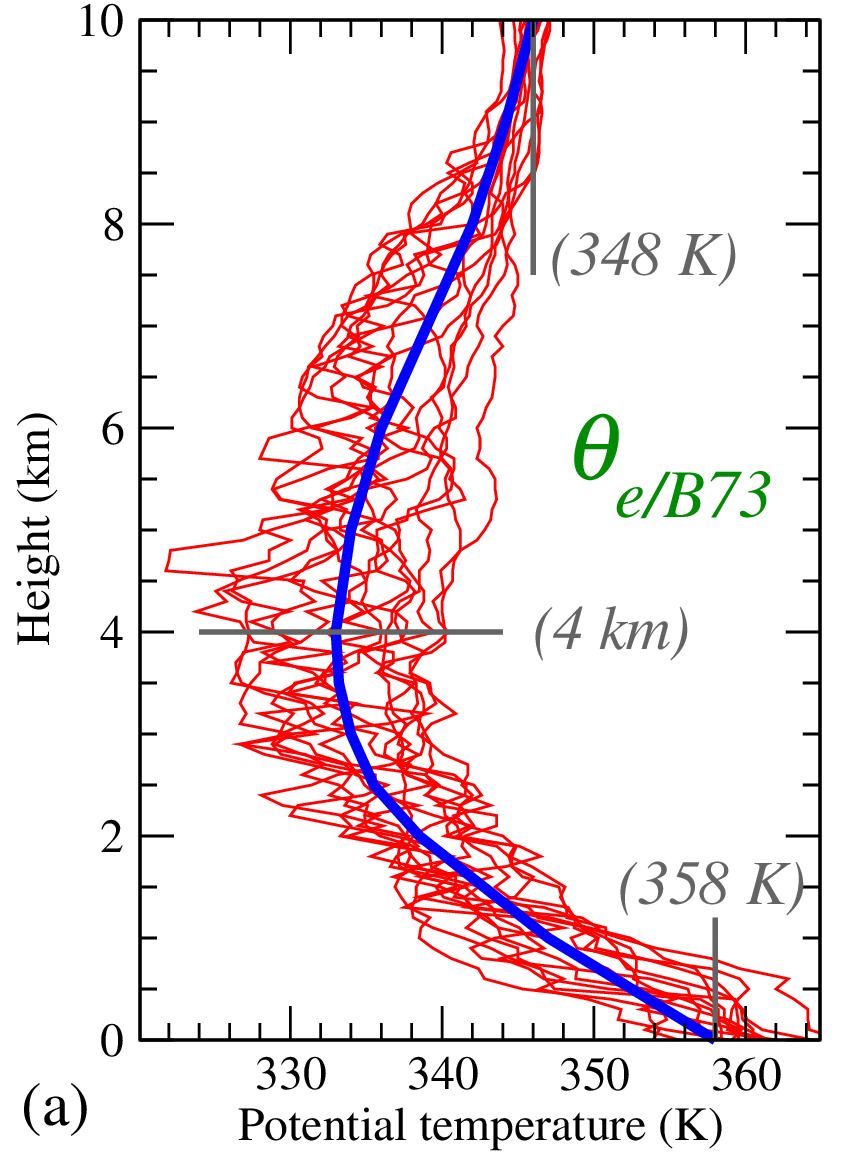}
\hspace{15mm}
\includegraphics[width=0.4\linewidth]{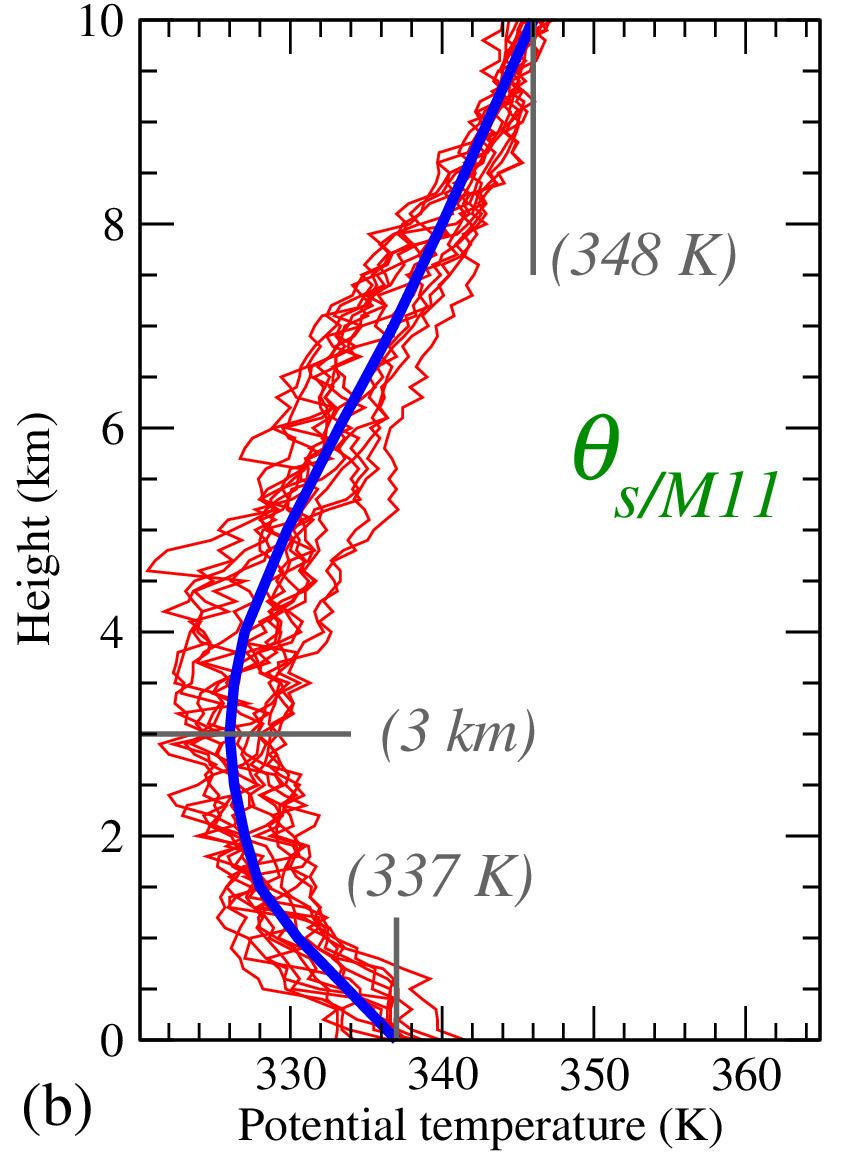}
}
\vspace*{-3mm} 
\caption{\it \small 
Vertical profiles for: (a) $\theta_{e/B73}$; and (b) $\theta_{s/M11}$ (K) for 14 dropwindsondes soundings into the Hurricane Earl on 1st of September 2010 (large-radii region, outside the eyewall).
Individual soundings are in
red lines
and the average profiles fitted by eye are in 
bold blue lines.
} 
\label{fig10}
\end{figure*}

Values of $\theta_{s/EM11}$ are conserved for the ascending branch between $1.5$ and $4$~km and for values close to $335$~K.
As in Fig.~\ref{fig8}, the locations of the minimum values of the streamlines are roughly aligned with the mean values of $\overline{\theta}_{s/M11}(z)$ (solid line).
Moreover, a dashed straight line has been added on Fig.~\ref{fig9} to show an intriguing linear organization from $2$ to $12$~km, which follows the location of all minimum values of $\Psi(\theta_{s/M11} ; z)$.
The upward and downward motions in the Hurricane DUMILE are located below and above this straight line and they correspond to monotonic increases and decreases in $\theta_{s/M11}$, respectively.
This simple linear organization is not observed for $\Psi(\theta_{e/E94} ; z)$ in Fig.~\ref{fig8}.


The underlying hypothesis justifying the study of the isentropic stream function is the fact that adiabatic and reversible processes lead to isentropic states, namely that any change in the moist-air entropy must correspond to diabatic sources or sinks.
Accordingly, the diabatic processes of mixing and entrainment are considered in MPZ to explain:
1) the decrease in $\theta_{e/E94}$ by following the streamlines of $\Psi(\theta_{e/E94} ; z)$ from the surface up to the melting level; and 
2) the increase in $\theta_{e/E94}$ above the melting level.
Clearly, if a similar increase of $\theta_{s/M11}$ is observed in Figs.~\ref{fig9} in dry regions above the freezing level, the decrease of $\theta_{s/M11}$ with $z$ below the freezing level does not exist, except for a small decrease close to the ground (below $0.5$~km) which is likely due to the impact of surface heating.

Comparing Figs.~\ref{fig8} and \ref{fig9} reveals that
$\theta_{e/E94} \: \approx \: \theta_l \; \exp(9 \: q_t)$ 
overestimates the impact of humidity in comparison with the third-law value 
$\theta_{s/M11} \: \approx \: \theta_l \; \exp(6 \: q_t)$.
The ratio $2/3$ for the coefficients in factor of $q_t$ in the exponential functions reveals new patterns and it is thus important to use the absolute entropy $s(\theta_{s/M11})$ when computing the isentropic stream function given by Eqs.~(\ref{eq_Psi_theta})-(\ref{eq_correl_rho_w}).

The differences between the potential temperatures $\theta_{e/E94}$ and $\theta_{s/M11}$ described by studying the outputs of the ALADIN model may depend on the various approximations that are made in all numerical models to describe moist-air processes and thermodynamic equations.

However, it is indicated in M11 that similarly large differences between $\theta_{e/E94}$ and $\theta_{s/M11}$ are observed for several vertical profiles of potential temperatures computed from the  radial flights in stratocumulus for the FIRE-I (First ISCCP Regional Experiment), ASTEX (Atlantic Stratocumulus transition Experiment), EPIC (East Pacific Investigation of Climate) and DYCOMS-II (DYnamics and Chemistry Of Marine Stratocumulus) campaigns.

Moreover, it is possible to validate the previous results derived from numerical outputs of ALADIN for the Hurricane DUMILE by comparing these results with several observed soundings into the Hurricane Earl \citep{Wang_al_BAMS_2015}.
Figs.~\ref{fig10}~(a) and (b) show that values of $\theta_{s/M11}$ near the ground are colder by about $11$~K than those at altitude at $400$~hPa or $10$~km height, whereas values of $\theta_{e/E94}$ are warmer by about $10$~K.
In addition, the height of minimum value and the dispersion of the profiles are both smaller with $\theta_{s/M11}$ than with $\theta_{e/E94}$.

The shape of individual and average profiles in Figs.~\ref{fig10}~(a) and (b) are thus similar to the descending parts of the stream functions in Figs.~\ref{fig8} and \ref{fig9} and of the loops of potential temperatures in Fig.~\ref{fig6}.
This confirms that the numerical outputs from the ALADIN model is a relevant laboratory to investigate isentropic processes, and that large differences do exist between the potential temperatures $\theta_{e/E94}$ and $\theta_{s/M11}$.

 \subsection{\underline{Impacts on heat input and work} 
 \underline{functions}} 
\label{subsection_impact_heat_work}


Many alternative definitions for the heat input or work functions have been considered in E86, E91, E94, \citet{Renno_Ingersoll_96,Pauluis_Held_2002_I,Pauluis_Held_2002_II}; and \citet{Goody_2003} and have been applied to different Carnot or steam cycles.
The aim of this section is to use the outputs of the Hurricane DUMILE to compare the numerical values of the heat input and work functions computed for several of these definitions.

According to \citet{degroot_mazur84} and for a state of local equilibrium, the Gibbs' differential equation for open systems can be written as
\begin{align}
 \!\! \!\!
  de_i + p \; d{\rho}^{-1} &
  = \:
  dh - {\rho}^{-1} \: dp \: 
 \: = \: 
    T \; ds
  \: + \sum_k \mu_k \; dq_k 
\label{eq_Gibbs} \: ,
\end{align}
where $e_i$ is the specific internal energy, $h$ the specific enthalpy,
$\rho$ the density and 
$\mu_k = h_k - T \: s_k$ the specific Gibbs' function for $k=0,1,2,3$ denoting dry-air, water vapor, liquid water and ice, respectively.

The work produced per unit mass of moist air $W = - \: \oint \: {\rho}^{-1} \: dp \:$ can then be computed by integrating Eq.~(\ref{eq_Gibbs}) along a closed circuit.
Since the integral of the total differential of enthalpy  $\oint \: dh$ vanishes along such a loop, $W$ can be written as
\begin{align}
 \!\! \!\!
 W & = \:
  - \: \oint \: \frac{dp}{\rho(p)} \: 
 \; = \: 
  \oint \: T(s) \; ds
  \: + \: \oint (\mu_v - \mu_d) \; dq_t 
\label{eq_Work} \: .
\end{align}
The last integral is derived from the last integral in Eq.~(\ref{eq_Gibbs}) for the steam cycle considered in Figs.~\ref{fig1} to \ref{fig4}, where $q_l=q_i=0$, $q_v=q_t$, and $q_d+q_t=1$, leading to $dq_l=dq_i=0$, $dq_d=-dq_v=-dq_t$, and thus to $\mu_d \: dq_d + \mu_v \: dq_v= (\mu_v - \mu_d) \; dq_t$.
The same equation for $W$ holds true if liquid or solid water exist, provided that changes of phases are reversible, which implies $(\mu_v-\mu_l)\:dq_l = (\mu_v-\mu_i)\:dq_i = 0$.





With the help of $h = \sum_k \: q_k \: h_k$, $s = \sum_k \: q_k \: s_k$ and $\mu_k = h_k - T \: s_k$, the Gibbs' equation (\ref{eq_Gibbs}) can be rewritten as
\begin{align}
 - \: {\rho}^{-1} \: dp \: & = \: 
  T \; \sum_k q_k \; ds_k 
  \: - \sum_k q_k \; dh_k 
\label{eq_Gibbs2} \: .
\end{align}
This relation clearly shows that, since $dh_k = c_{pk} \: dT$ (for all species), $ds_k = c_{pk} \: dT/T - R_k \: dp_k/p_k$ (for gases) and $ds_k = c_{pk} \: dT/T$ (for liquids or solids), the reference values of enthalpies and entropies have no impact and may remain undetermined if the aim is to compute $- \: dp / \rho$.

Accordingly, the reference enthalpies $h_{v0}$ and $h_{d0}$ can be discarded in the computations of the enthalpies of water vapor and dry air contained in the last integral of Eq.~(\ref{eq_Work}) which depends on 
$(\mu_v - \mu_d) = (h_v - h_d) - T \: (s_v - s_d)$.
This property holds true because 
$\oint \: (h_{v0} - h_{d0} ) \: dq_t = (h_{v0} - h_{d0} ) \: \oint \: dq_t = 0$ since 
$\oint \: dq_t \equiv 0$ for a closed loop. 

In contrast, the reference entropies must be taken into account in the computations of 
$\oint \: T \: (s_{d0} - s_{v0} ) \: dq_t = (s_{d0} - s_{v0} ) \: \oint \: T \: dq_t$, because 
$\oint \: T(q_t) \: dq_t \neq 0$ if the area of the loop is different from zero in a $T$-$q_t$ diagram.
However, the impact of the reference entropies $s_{d0}$ and $s_{v0}$ on the last integral in Eq.~(\ref{eq_Work}) must cancel out with the impact on the integral $\oint \: T(s) \; ds$.
This means that a change in the reference entropies must modify the values of the two integrals on the right hand side of Eq.~(\ref{eq_Work}), although their sum, which is equal to $W = - \: \oint \: dp \: / \rho$, must not depend on $s_{d0}$ or $s_{v0}$.


\begin{table*}
\caption{\it \small 
The heat inputs $W_H$ (or $W_{H(d)}$) in J~kg${}^{-1}$ computed for the six moist-air entropies considered in section~\ref{subsection_s}.
Here $W_H$ (or $W_{H(d)}$) represents the areas of the six loops shown in Fig.~\ref{fig7} (Riemann integrals).
The wind scale $V = \sqrt{2\,W_H}$ is in units of m~s${}^{-1}$.
\label{Table4}
}
\centering
\vspace*{2mm}
\begin{tabular}{cccccccc}
\hline
``$\theta$'' & 
$\theta_{es/E86}$ & $\theta_{s/M11}$ & $\theta_{e/MPZ}$ & $\theta_{s}$ & $\theta_{es/E86}$  & $\theta_{e/E94}$  \\ 
\hline
``$s$'' 
& $s/q_d$ & $s$ & $s/q_d$ & $s/q_d$ & $s/q_d$ & $s/q_d$ \\ 
``$c_{p}$''
& $c_{pd}$ & $c_{pd}$  & $c^{\ast}_{pl}$ & $c_{pd}$ & $c^{\ast}_{pl}$ & $c^{\ast}_{pl}$ \\ 
 \!\!\!\! Offset \!\!\!\!
& $\!\!-5650\!\!$ & $\!\!-6850\!\!$ & $+550$ & $\!\!-6700\!\!$ & $\!\!-5610\!\!$ & $\!\!-2300\!\!$\\ 
 ``$W_H$'' \!\!\!
& $442$ & $870$ & $1158$ & $1528$ & $2758$ & $3405$\\ 
 \!\!\! ``$V$'' \!\!\! 
& $29.7$ & $41.7$ & $48.1$ & $55.3$ & $74.3$ & $82.5$\\ 
\hline
\end{tabular}
\vspace*{-4mm} 
\end{table*}

The integrals in Eq.~(\ref{eq_Work}) are computed in E94, \cite{Goody_2003} and P11 with all extensive quantities expressed per unit of dry air, and thus with the specific values of $h$ and $s$ replaced by $h/q_d$ and $s/q_d$.
Furthermore, the reference entropies for dry air and liquid water are assumed to vanish at $T_0 = 0$~C in P11, leading to 
\begin{align}
 \!\! \!\! \!\! \!
 W_{(d)} & =  - \:
  \oint  \: \frac{dp}{\rho_d(p)}
 \: = \:  \oint \: T \; d\left( \frac{s}{q_d} \right)
  \: + \:
  \oint \: g_v \; dr_t 
\label{eq_Work_P11} , \\
 \!\! \!\! \!\! \!
 g_v & =  c_l \left[ \: T - T_0 - T \: \ln(T/T_0) \: \right]
 \: + \: R_v \: T \: \ln(H_l)
\label{eq_gv_P11} .
\end{align}
The work functions $W$ or $W_{(d)}$ are approximated in the older studies on Carnot cycles by considering only the first integral on the right-hand sides of Eqs.~(\ref{eq_Work}) or (\ref{eq_Work_P11}), see E91 (Eqs.~4 and 10) and \citet[Eq.~3]{Renno_Ingersoll_96}, yielding
\begin{align}
W \; \approx \; W_H      & \: = \; \oint \: T(s) \: \; ds
\label{eq_W1}  \; \; \; \; \mbox{and} \\
W_d \; \approx \;W_{H(d)} & \: = \; \oint \; T\!\left( \frac{s}{q_d} \right) \; \: d\!\left( \frac{s}{q_d} \right)
\label{eq_W1d} \: .
\end{align}
$W_H$ and $W_{H(d)}$ are commonly called ``heat input''.
They are equal to the areas enclosed by the loops in the temperature-entropy diagram of Fig.~\ref{fig7} and they are independent of the global offset chosen for each loop.
Values of the heat input $W_H$ and $W_{H(d)}$ listed in Table~\ref{Table4} are calculated using the simple but robust first-order Riemann method.

Large factors, of $0.5$ and $4$, are observed between $W_H \approx 870$~J~kg${}^{-1}$, computed with the third-law value $s(\theta_s)$, and $W_{H(d)} \approx 442$~J~kg${}^{-1}$ or $W_{H(d)} \approx 3405$~J~kg${}^{-1}$, computed with $s(\theta_{es/E86})/q_d$ or $s(\theta_{e/E94})/q_d$, respectively.
Furthermore, the impact of $c^{\ast}_{pl}$ is very large compared to that of $c_{pd}$, leading to a factor of more than $6$ for $W_{H(d)}$ computed with the same potential temperature $\theta_{es/E86}$ ($442$ versus $2758$~J~kg${}^{-1}$).

The impact on ``$W_H$'' of the definition ``per unit of dry air'' versus the definition ``per unit of moist air'' (specific value) can be evaluated by comparing $s$ and $s/q_d$ for the same third-law value $\theta_{s/M11}$.
The impact $1528 - 870 = 658$~J~kg${}^{-1}$ is large, leading to an increase of more than $75$~\% for the definition ``per unit of dry air''.

The impact of the choice of the different formulations for the moist-air entropy can be evaluated differently, by computing the wind scale $W_H = V^2/2$, with $V$ becoming a crude proxy for the surface wind that a perfect Carnot engine might produce.
The last line in Table~\ref{Table4} shows that $V$ would vary from about $30$ to more than $80$~m~s${}^{-1}$.

Since temperatures can vary by $30$~K along the Carnot cycle, and for an accuracy of about $1$~J~K${}^{-1}$~kg${}^{-1}$ for entropies, errors in computations of $W_H$ and $W_{H(d)}$ are of the order of $30$~J~kg${}^{-1}$.
This means that the observed differences, of the order of hundreds or thousands of J~kg${}^{-1}$, are significant and, since the moist-air entropy is a state variable, the third-law formulation computed with $\theta_{s/M11}$ must be used to compute both $W_H$ and $W_{H(d)}$.

Independent computation of the integrals on the right-hand sides of Eqs.~(\ref{eq_Work}) and (\ref{eq_Work_P11}) is thus impossible without knowledge of the third-law reference entropies, because only the sum of these integrals is independent of these reference values.
It is thus necessary to compute the full work function and to switch from a Carnot cycle to a steam cycle, for which changes in water vapor have a large impact via the last integrals of Eqs.~(\ref{eq_Work}) and (\ref{eq_Work_P11}).

\begin{figure}[h]
\centerline{\includegraphics[width=0.99\linewidth]{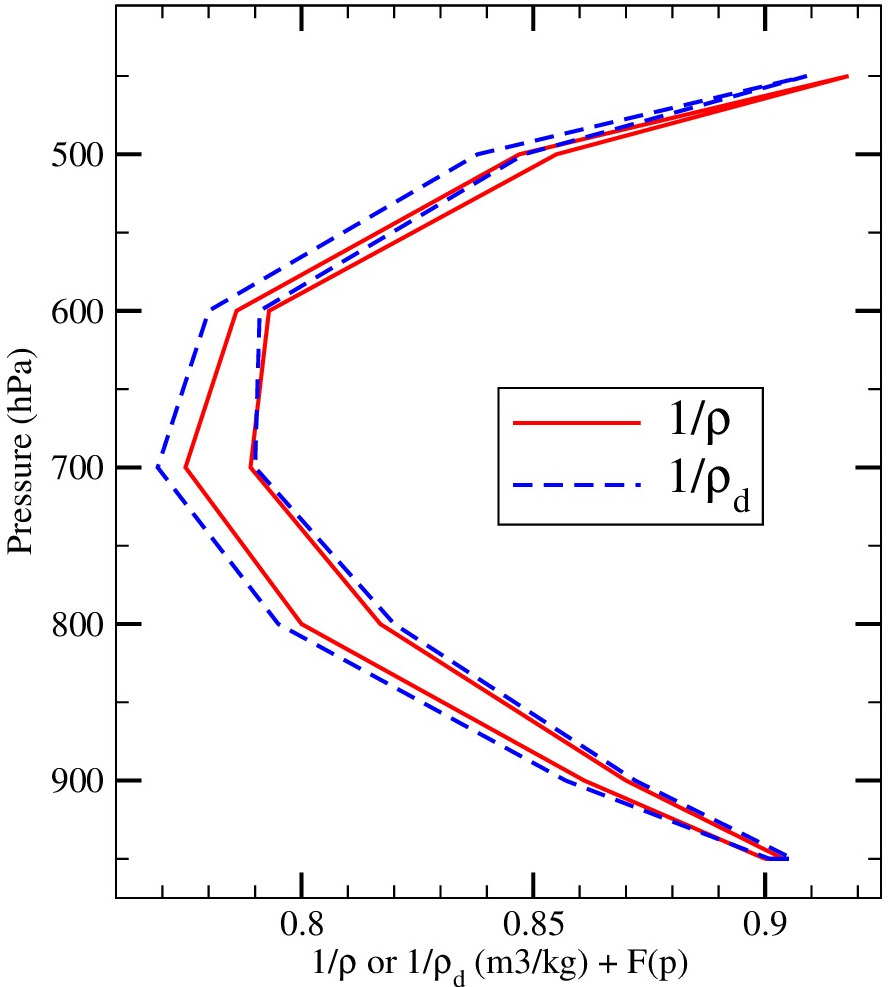}}
\vspace*{-3mm} 
\caption{\it \small 
Vertical profiles of $1/\rho$ 
(solid red line)
and $1/\rho_d$ 
(dashed blue line)
 for the $15$ points marked in Figs.~\ref{fig1}-\ref{fig4}.
Units are kg~m${}^{-3}$.
The skew diagram was built by adding a shift depending on a function $F(p)$ described in the text.
} 
\label{fig11}
\end{figure}

\begin{table*}
\caption{\it \small 
Values of $W$ and $W_{(d)}$ computed by Eqs.~(\ref{eq_Work}) and (\ref{eq_Work_P11}) with Riemann integrals.
\label{Table5}
}
\centering
\vspace*{2mm}
\begin{tabular}{cccc}
\hline
$\!\!W / M11 \!\!$       & $= \: \oint T \: ds$ &       $\!\!+ \: \oint (\mu_v-\mu_d) \: dq_v\!\!$ & $+ \: \Delta$  \\ 
$455$ & $870$ & $-371$ & \!\!$-44$ $(\approx 9.7 \: \%)$\!\! \\
\hline
$\!\!W_{(d)} / P11 \!\!$ &\!\! $= \: \oint T \: d(s/q_d)$ \!\!& $+ \: \oint g_v \: dr_v$ \!\!& $+ \: \Delta$  \\ 
$703$ & $1201$ & $-458$ & \!\!$-40$ $(\approx 5.7 \: \%)$\!\!\\
\hline
\end{tabular}
\vspace*{-4mm} 
\end{table*}


It is assumed in E88 that part of the energy available from the steam cycle is actually used to lift water and is not available for the generation of kinetic energy.
However, insofar as Eqs.~(\ref{eq_Work}) and (\ref{eq_Work_P11}) are considered without including these possible effects,
the values of all integrals appearing 
for $W$ and $W_{(d)}$ are listed in Table~\ref{Table5}.
The left and right hand sides of either $W$ or $W_{(d)}$ are equal within an accuracy of about $40$~J~kg${}^{-1}$, which is better than $10$~\% for $W$ and $6$~\% for $W_{(d)}$.
This accuracy can be improved by using a Simpson second-order method, leading to $2.5$~\% for $W$ and $0.5$~\% for $W_{(d)}$ (not shown).
This confirms the theoretical methods for establishing both Eqs.~ (\ref{eq_Work}) and (\ref{eq_Work_P11}).
In particular, it proves that the reference enthalpies and entropies of dry air and water vapor do not have an impact on $W$ and $W_{(d)}$.





It seems somewhat surprising that $W_{(d)}$ can be $(703-455)/455 = 55$~\% higher than $W$!
However, this large difference is not a numerical artifact; it merely comes from the replacement of $\rho$ by $\rho_d$ in the computation of $W_{(d)}$.
To demonstrate this result, values of $\rho$ and $\rho_d$ are shown in Fig.~\ref{fig11} for the $15$ points of the steam cycle of the Hurricane DUMILE.
To facilitate visual comparisons, the terms $F(p) = 0.7 \: (p-950)/450$ and $F(p) - 0.014$ have been added to $1/\rho$ and $1/\rho_d$, respectively.
These terms are the same for a given pressure and they do not affect the computation of the differences between upward and downward values.

Although the replacement of $1/\rho$ by $1/\rho_d$ leads to a relative change of less than $2$~\% if $q_v<20$~g~kg${}^{-1}$, the differences between upward and downward values are much larger: $(1/\rho_d)_{up}-(1/\rho_d)_{down}$ is more than $40$~\% larger than $(1/\rho)_{up}-(1/\rho)_{down}$ for all levels between $900$ and $600$~hPa.
Therefore, the area of the loop $W_{(d)}$ is logically about $45$~\% larger than $W$, and the mixing-ratio view expressed per unit mass of dry air leads to a value $W_{(d)}$ that largely overestimates the work function $W$ defined in thermodynamics with specific values.
The consequence is the need to consider $W$ for studying steam cycles, since the use of $W_{(d)}$ would lead to large errors and to erroneous physical meanings if $W_{(d)}$ was to be converted into potential or kinetic energy.

The aim of the present paper is not to provide the most general formulation for the work functions $W$ given by Eqs.~(\ref{eq_Work}) or~(\ref{eq_Work_P11}).
Indeed, $W$ may be subject to some uncertainty with, for example, some additional terms which are added to Eq.~(\ref{eq_Work_P11}) in E88.
These additional terms are intended to take into account possible irreversible entropy sources ranging from the fallout of condensed phase water to irreversible mixing across water concentration gradients to freezing of supercooled water.
Moreover, it is assumed in E88 that part of the energy available from the steam cycle is actually used to lift water and is not available for the generation of kinetic energy.

Differently, it is shown in the present paper that once a formulation for $W$ is chosen, with or without the additional terms of E88, and provided that it is indeed derived from the Gibbs equation, the value of the work function may depend on the reference entropies of dry air and water vapor.
And in this case, it is essential to use the values derived from the third law of thermodynamics.

 \subsection{\underline{Impacts on the efficiency of steam}
 \underline{cycles}} 
\label{subsection_impact_efficiency}

It was shown in the previous section that it is impossible to compute the integrals on the right-hand sides of Eqs.~(\ref{eq_Work}) and (\ref{eq_Work_P11}) independently.
Nevertheless, calculations of this kind are often used in atmospheric science to evaluate the efficiency factor of a thermal engine.

The efficiency $E$ of a tropical cyclone considered as a thermal engine was defined in \citet{Palmen_Riehl_1957} as ``the ratio of the mechanical energy produced to the heat released'' and $E$ was evaluated to be about $2$ to $3$~\%.

The efficiency of a hurricane considered as a steam cycle is defined in P11 by taking account of the last integrals in Eqs.~(\ref{eq_Work}) and (\ref{eq_Work_P11}), which are intended to model the effects due to the open cycle in which mass is added and removed.
These Eqs.~(\ref{eq_Work}) and (\ref{eq_Work_P11}) can be rewritten as $W = Q_1 = Q_2$ and $W_{(d)} = Q_{1(d)} = Q_{2(d)}$, where
\begin{align}
 \! \! \!
 Q_1 & = \:
  \oint \:\delta Q_1
  \; = \; \oint \: \left( 
     dh \: - \: \frac{dp}{\rho} 
          \right)  \; \; \; \; \mbox{and}
\label{eq_deltaQ1} \\
 \! \! \!
 Q_2 & = \:
  \oint \:\delta Q_2
  \; = \; \oint \: \left[ \: \:
     T \; ds \: + \: \left( \mu_v - \mu_d \right) \; dq_t 
          \: \: \right] \: 
\label{eq_deltaQ2} \: ,
\end{align}
and
\begin{align}
 \! \! \! \! \!
 Q_{1(d)} & = \:
  \oint \:\delta Q_{1(d)}
  \; = 
       \oint \: \left[ \:
     d\left(\frac{h}{q_d}\right)
  \: - \: \frac{dp}{\rho_d} \:  
          \: \right]
\label{eq_deltaQ1d}  \; \;  \mbox{and} \\
 \! \! \! \! \!
 Q_{2(d)} & = \: 
  \oint \:\delta Q_{2(d)}
  \; = 
  \oint \: \left[ \: 
     T \; d\left(\frac{s}{q_d}\right)
  \: + \: g_v \; dr_t  
          \: \right] 
\label{eq_deltaQ2d} \: .
\end{align}
The ``energy input'' is defined in P11 as
$Q_{in} = \oint \: \delta Q^{+}$, where $\delta Q^{+} = \mbox{max}(0, \delta Q)$ is different from zero only for positive values of $\delta Q$.
The mechanical efficiency is then defined and computed in P11 by the ratio 
\begin{align}
\chi & \: = \; \frac{Q}{Q_{in}}
\label{eq_eta} \: ,
\end{align}
where all quantities are expressed per unit mass of dry air.
From Eq.~(22) in P11, this ratio can be approximated by
$\chi_S \approx (T_{in}- T_{out})/1540 \:\mbox{K}$, where $T_{in}$ is the highest and $T_{out}$ the lowest temperature of the cycle and where $(R_d \: L_v) / (R_v \ c_{pd}) \approx 1540$~K.
The steam cycle considered for the Hurricane DUMILE corresponds to $T_{in} \approx 296$~K and $T_{out} \approx 265$~K, leading to $\chi_S \approx 0.020$.

Table~\ref{Table6} shows that $W = Q_1 \approx Q_2$ and $W_{d} = Q_{1(d)} \approx Q_{2(d)}$ with the same accuracy of about $40$~J~kg${}^{-1}$ previously noted for the Riemann method.
The differences $Q_1 \neq Q_{1(d)}$ and $Q_2 \neq Q_{2(d)}$ are due to the use of mixing ratios and of quantities expressed per unit mass of dry air in P11.
This means that the way the work function is defined and computed may largely impact the diagnosis of the external heating required with any transformation of moist air, and may modify the numerator of the efficiency factor given by Eq.~(\ref{eq_eta}).

Similarly, the way $\delta Q^{+}$ and $Q_{in}$ are computed may impact the denominator of Eq.~(\ref{eq_eta}).
The computations of $\delta Q^{+}$ for $Q_1$, $Q_2$, $Q_{1(d)}$ or $Q_{2(d)}$ are especially uneasy because they rely on accurate definitions of all the enthalpies ($h$ of $h/q_d$), the entropies ($s$ or $s/q_d$) and the Gibbs functions ($\mu_v$, $\mu_d$ or $g_v$).
Moreover, whereas $\oint \: dh = 0$ for a closed cycle, the integral of $dh$ is different from zero if it is computed only for those points where $\delta Q_1 > 0$.

The enthalpy ($h/q_d$), entropy ($s/q_d$) and Gibbs functions ($g_v$) appearing in Eqs.~(\ref{eq_deltaQ1d}) and (\ref{eq_deltaQ2d}) are those defined in P11.
The specific entropy involved in Eqs.~(\ref{eq_deltaQ2}) is the third-law value defined in M11 and recalled in Eq.~(\ref{eq_s_thetas_M11}).
The specific enthalpy in Eqs.~(\ref{eq_deltaQ1}) is the one defined in \citet{Marquet15c,Marquet15a}:
\vspace*{-1mm}
\begin{align}
\! \! \! \! \!
  h & \: = \; h_{\rm ref} 
      \: + \: c_{pd} \: T 
      \: + \: L_h \: q_t 
      \: - \: L_v \: q_l 
      \: - \: L_s \: q_i
 \label{eq_h}   \: , \\
 \! \! \! \! \!
  L_h & \: = \; h_v - h_d 
  \: = \; L_h(T_r) 
  \: +  (c_{pv}-c_{pd}) \: (T-T_r)
 \label{eq_Lh2} \: ,
\end{align}
where
$L_h(T_r)  \approx  2603  \mbox{~kJ} \mbox{~kg}^{-1}$ and
$h_{\rm ref} \approx 256$~kJ~kg${}^{-1}$.
The Gibbs functions $\mu_v$ and $\mu_d$ are computed by $h_v - T \: s_v$ and $h_d - T \: s_d$, 
where $h_d(T_r)  \approx 530 \mbox{~kJ} \mbox{~kg}^{-1}$ and $h_v(T_r)\approx  3133  \mbox{~kJ} \mbox{~kg}^{-1}$.

The latent heat $L_h(T)$ defined in Eq.~(\ref{eq_Lh2}) depends on the values of $L_h(T_r)$, in the same way as $L_v$ depends on $L_v(T_r)  \approx  2501  \mbox{~kJ} \mbox{~kg}^{-1}$, and in the same way as  the entropy potential temperature ${\theta}_{s/M11}$ defined in Eq.~(\ref{eq_thetas}) depends on the third-law values  $\Lambda_r$, ${(s_v)}_r$ and ${(s_d)}_r$.

However, $L_h(T)$ is independent of the reference temperature $T_r$ because $L_h(T_r) + (c_{pv}-c_{pd}) \: T_r$ is a constant, in the same way as $L_v$ and $L_s$ are independent of the value $T_r$ used to compute them, and in the same way as ${\theta}_{s/M11}$ does not depend on $T_r$, $p_r$ and $r_r = r_{sw}(T_r,p_r) =  \varepsilon \: e_r \, / \,  (p_r - e_r)$ provided that $e_r = e_{sw}(T_r)$.

\begin{table}
\caption{\it \small 
External heating ($Q$), energy inputs ($Q_{in}$) and mechanical efficiency ($\chi$).
All integrals are computed by using the first-order Riemann method.
\label{Table6}
}
\centering
\vspace*{2mm}
\begin{tabular}{ccc}
\hline
$Q_{1}     =  455 $ & $Q_{in/1} = 22114$ & $\chi_{1} = 0.021$ \\
$Q_{2}     =  499 $ & $Q_{in/2} = 22150$ & $\chi_{2} = 0.023$ \\
\hline
$Q_{1(d)}     =  703 $ & $Q_{in/1(d)} = 21833$ & $\chi_{1(d)} = 0.033$ \\
$Q_{2(d)}     =  743 $ & $Q_{in/2(d)} = 21871$ & $\chi_{2(d)} = 0.035$ \\
\hline
\end{tabular}
\vspace*{-4mm} 
\end{table}

The efficiency factors shown in the last column of Table~\ref{Table6} are almost the same for $Q_{1}$ and $Q_{2}$ on the one hand, and for $Q_{1(d)}$ and $Q_{2(d)}$ on the other hand.
However, the values, respectively close to $2.2$~\% and $3.4$~\%, are not compatible.
Although these two efficiency values are close to those derived in \citet{Palmen_Riehl_1957}, only the former value is close to the approximate value $\chi_S \approx 0.020$ derived in P11.
The explanation for this important difference seems to come from the energy input terms $Q_{in/1}$ and $Q_{in/2}$, which are only $1$~\% smaller than $Q_{in/1(d)}$ and $Q_{in/2(d)}$, whereas the full work functions $Q_{1(d)}$ and $Q_{2(d)}$ are $50$~\% larger than $Q_{1}$ and $Q_{2}$.

The definitions of \citet{Palmen_Riehl_1957} or P11 for the efficiency factor may not be the most relevant ones.
It is thus probably necessary either to change these definitions 
in order to be independent in the choice of the reference values of entropies and enthalpies, or to use the third-law potential temperature, with all specific quantities expressed per unit of moist air, in order to comply with the recommendations of open-system thermodynamics.

 \section{\underline{Discussion and Conclusion}} 
\label{section_conclusions}

Isentropic analysis is probably a powerful tool for investigating moist-air energetics by plotting moist-air isentropes or by computing isentropic mass fluxes.
However, the quality and the realism of such an analysis rely on a clear definition of the moist-air entropy.

It has been shown, here, that the way the potential temperatures $\theta_s$, $\theta_e$ or $\theta_{es}$ are defined as ``equivalents'' of the moist-air entropy significantly impacts the computations and plots of isentropic surfaces, making ``isentropic'' analyses similar to those published in MPZ or P11 uncertain.

It has been pointed out that the stream function, the heat input and the efficiency factor may be largely modified by the way in which the entropy and the enthalpy are defined: $s$ or $s/q_d$, $h$ or $h/q_d$; modified reference values for entropies and enthalpies; the choice of $\theta_s$, $\theta_e$ or $\theta_{es}$, the choice of $c_{pd}$ or $c^{\ast}_{pl}$ as a factor of the logarithm; {\it etc\/}.

The issue associated with the ``per unit mass of dry air'' view can be understood as follows.
If isentropic processes are defined as in E94, MPZ and P11, with constant values of $s/q_d$, the definitions of the geopotential, the wind components or the kinetic energy one should modify accordingly by plotting, for instance, $g\: z / q_d$, $u/q_d$, $v/q_d$ or $(u^2+v^2)/(2 \: q_d)$.
These definitions are unusual in the thermodynamics of open systems.
Moreover, if $s/q_d$ could be defined within a global constant $C$, the specific value $s$ would depend on $q_d \: C$, which varies with $q_d$ and renders the integral $S = \iiint s \: \rho \: d\tau$ indeterminate, because it would depend on $\iiint q_d \: C \: \rho \: d\tau$, where $C$ is an unknown term.
This is not realistic.


The third-law of thermodynamics, which can be based on calorimetric, quantum or statistical physics, can now be considered as well-established.
It provides a fortunate opportunity to compute, describe, analyze and possibly understand several features of atmospheric energetics in a new way.

It is important to clearly differentiate the wish to define conserved variables like $\theta_l$, $\theta'_w$, $\theta_e$ or $\theta_{es}$, on the one hand, from the need to define the specific moist-air entropy by computing the third-law value $\theta_s$, on the other hand.

These two aspects are complementary because a given process which might correspond to the law of conservation of $\theta_e$, for instance, may correspond to a change of entropy which can be precisely computed by using the third law value $s(\theta_s)$.
Conversely, a non-trivial isentropic process where $\theta_s$ is conserved, but where both $\theta_l$ and $q_t$ vary with time and space, might not correspond to a conservation of any of $\theta'_w$, $\theta_e$ or $\theta_{es}$.

Moist-air isentropic surfaces are not subject to uncertainty in the natural world and it is likely that the third-law definitions $\theta_s$ and $s(\theta_s)$ given by Eqs.~(\ref{eq_thetas}) and (\ref{eq_s_thetas_M11}) are the most relevant, since they are based on general thermodynamic principles and use specific values expressed per unit mass of moist air, as with all other variables in fluid dynamics.





\vspace{4mm}
\noindent
{\large \bf \underline{acknowledgments}}
\vspace{2mm}

I thank the two reviewers for their comments, which helped to improve the manuscript.
The data for the Hurricane Earl was provided by NCAR/EOL under the sponsorship of the National Science Foundation. \url{https://data.eol.ucar.edu/}
\vspace{2mm}

\vspace{4mm}
\noindent
{\bf \underline{Appendix A. The history of the third law of}
     \underline{thermodynamics.}}
             \label{appendixA}
\renewcommand{\theequation}{A.\arabic{equation}}
  \renewcommand{\thefigure}{A.\arabic{figure}}
   \renewcommand{\thetable}{A.\arabic{table}}
      \setcounter{equation}{0}
        \setcounter{figure}{0}
         \setcounter{table}{0}
\vspace{2mm}


It is surprising that so many potential temperatures may serve to compute the moist-air entropy at the same time. 
Since the moist-air entropy is a state function, the status of the isentropic or diabatic evolution of a given parcel of moist air is an observable feature and, in the words of \citet[ p.157]{Richardson22}: ``approximations are not here permissible''.
The only way to solve this problem is to rely on general thermodynamics where the aim of the third law is precisely to provide an absolute and non-ambiguous definition of the entropy state function.
According to reviews of the old treatises and papers of thermodynamics \citep{Lewis_Randall_1923,Wilks61,Barkan_1999,Coffey_2006,Klimenko12}, there are three ways to express the third law.

In the mid and late 19th century, an important problem in chemistry was how to improve the understanding of chemical reactions and to predict their spontaneity.
The answer given by \citet{Gibbs_1875_1878,Gibbs_1878art} was to compute the sign of the Gibbs function (or free enthalpy) of reaction 
\vspace*{-1mm}
\begin{align}
  \Delta G & \: = \; \Delta H \: - \: T \: \Delta S \: = \: - \; T \; (\Delta S)_{\rm \, tot}
\label{eq_G_H_TS_Gibbs} \: ,
\end{align}
where $\Delta H$ and $\Delta S$ are the enthalpy and entropy of reaction and $T$ is the temperature.
Negative values of $\Delta G$ would make the reaction proceed spontaneously because, according to the second law of thermodynamics, the total entropy of the system and its surroundings, $(\Delta S)_{\rm \, tot}$, would increase.

\citet{Gibbs_1875_1878} explicitly introduced two constants of integration appearing in the computation of the specific energy, entropy, free energy and free enthalpy for an ideal gas (pages 150 to 152).
He then arrived at the definitions of the entropy, free energy and free enthalpy of a mixture of ideal gases (equations 278, 279 and 293 pages 156 and 163).
These definitions depend on the constants of integration, which, {\it a priori\/}, are different for each individual gas and are multiplied by the individual variable concentrations.
Moreover, since $\Delta S$ is multiplied by the variable temperature in the computation of $\Delta G$ given by Eq.~(\ref{eq_G_H_TS_Gibbs}), the free enthalpy computed in equation 293 of \citet{Gibbs_1875_1878} must depend on the constants of integration for the specific entropies.

The problem of finding the numerical values of $\Delta H$, $\Delta S$ and $\Delta G$ was next considered by \citet{Chatelier_1888,Lewis_1899,Richards_1902,vant_Hoff_1904}; and \citet{Haber_1905} with an alternative version of Eq.~(\ref{eq_G_H_TS_Gibbs}).
They all focused on the free energy $A = \Delta G$ on both sides of Eq.~(\ref{eq_G_H_TS_Gibbs}) by using the Helmholtz equations
\vspace*{-1mm}
\begin{align}
  A & \: = \Delta H \: + \: T \; \frac{dA}{dT} 
\; \; \; \; \mbox{or} \;  \; \;
  \frac{d}{dT}\left( \frac{A}{T} \right) 
     = \:
   - \: \frac{\Delta H}{T^2}
\label{eq_A_Nernst_1906} \: ,
\end{align}
where $A$ represents the maximum available work (``maximaler Arbeit'').
Comparison of Eqs.~(\ref{eq_G_H_TS_Gibbs}) and (\ref{eq_A_Nernst_1906}) shows that the derivative at constant pressure ${dA}/{dT}$ is equal to $-\:\Delta S$ (the opposite of the entropy of reaction).

The advantage of Eq.~(\ref{eq_A_Nernst_1906}) is that it provides the actual possibility of measuring the heat of reaction $\Delta H(T)$ for different absolute temperatures from laboratory experiments, and then of finding the value of $A(T) / T$ by an integration of Eq.~(\ref{eq_A_Nernst_1906}) between $T_0$ and $T$, leading to
\vspace*{-1mm}
\begin{align}
  A(T) & 
   = 
     - \; T \int^{T}_{T_0} \frac{\Delta H(T')}{(T')^2} \: dT'
 \;  + \; T  \left[ \: \frac{A(T_0)}{T_0} \: \right]
\label{eq_A_Nernst_1906_int}  .
\end{align}

The knowledge of $A(T)$ from Eq.~(\ref{eq_A_Nernst_1906_int}) should allow the spontaneity of any chemical reaction to be predicted, for a given temperature $T$; it depends on the sign of $A(T) = \Delta G$.
However, the constant of integration ${A(T_0)}/{T_0}$ is multiplied by $T$ and influences the value of $A(T)$, and thus possibly the sign of $\Delta G(T)$.
This cannot be true because it would bring some arbitrariness into the spontaneity of the chemical reaction, making it possible to modify at will the behavior of chemical reactions like
$\mbox{H}_2 + \mbox{O}_2 = \mbox{H}_2\mbox{O}$
in the atmosphere by choosing any arbitrary value for the constant of integration ${A(T_0)}/{T_0}$ for the temperature $T_0$.

\begin{table*}
\caption{\it \small The first three columns list the standard entropies for five atmospheric gases at $298.15$~K and $p_0 = 1000$~hPa published in \citet[C98]{Chase_98} and \citet[GR96]{Gokcen_Reddy_1996}.
The statistical physics method is based on $S = k \: \ln(W) + S_0$ with $S_0=0$.
The calorimetric method is based on (\ref{eq_s_third_law}) with the third law 
$S( \, 0 \: \mbox{K}, p_0)=0$.
The fourth column corresponds to values computed with $S( \, 0 \: \mbox{K}, p_0)=0$ and the calorimetric datasets  described in \citet{Marquet15a}.
Units are cal~K${}^{-1}$~mol${}^{-1}$ throughout.}
\centering
\vspace*{2mm}
\begin{tabular}{ccccc}
\hline
\!\!          \!\! &\!\!  C98/Sat.  \!\! & \!\!\! GR96/Stat. \!\!\! &\!\!\! GR96/Cal.\!\!\! & \!\!\! M15/Cal. \!\!\! \\
\hline
\!\! Ar       \!\! &\!\!  $ 37.000 \: \pm 0.001$ \!\! & $37.00$ &\!\!\! $36.96 \: \pm 0.2$ \!\!\! & \!\!\!  \!\!\!  \\
\!\! O${}_2$  \!\! &\!\!  $ 49.031 \: \pm 0.008$ \!\! & $49.02$ &\!\!\! $49.12 \: \pm 0.1$ \!\!\! & \!\!\! $49.7 \: \pm 0.4$ \!\!\!  \\
\!\! N${}_2$  \!\! &\!\!  $ 45.796 \: \pm 0.005$ \!\! & $45.78$ &\!\!\! $45.94 \: \pm 0.2$ \!\!\! & \!\!\! $46.0 \: \pm 0.2$ \!\!\!  \\
\!\! H${}_2$O \!\! &\!\!  $ 45.132 \: \pm 0.010$ \!\! & $45.12$ &\!\!\! $44.31  	 $       \!\!\! & \!\!\! $45.2 \:  \pm 0.1$ \!\!\!  \\
\!\! CO${}_2$ \!\! &\!\!  $ 51.098 \: \pm 0.029$ \!\! & $51.09$ &\!\!\! $51.13 \: \pm 0.1$ \!\!\! & \!\!\!  \!\!\!  \\
\hline
\end{tabular}
\vspace*{-4mm}
\end{table*}

This problem was solved by \citet{Nernst_1906} by assuming (page 49) a ``new hypothesis'' which is nowadays called the ``Heat Theorem'':
\vspace*{-1mm}
\begin{equation}
\! \! \! \! \!
  \lim \: \frac{dA}{dT} 
  \: = \: 
  \lim \: \frac{d(\Delta H)}{dT} 
  \: = \:  0 
  \; \; \mbox{when} \; \; \; T = 0 \; \mbox{K} \; . 
\label{eq_Heat_Theorem_Nernst_1906}
\end{equation}

\noindent The heat theorem was not expressed in terms of the entropy but, instead, was based on the Helmholtz function $A$, leading to 
$A(T_0) / \: T_0 \rightarrow \: 0$ as $T_0  \rightarrow \: 0$~K 
in Eq.~(\ref{eq_A_Nernst_1906_int}).
This is the third law of thermodynamics in its original formulation suggested by Nernst.

Nernst tested the consequences of his theorem by computing the theoretical values of the affinities or the reaction rates of several chemical reactions.
The good agreement he showed with experimental results supported his theorem.
The hope of Nernst was then to demonstrate his theorem starting from the observed anomalous behavior of specific heats, which tend to zero at low temperature and with $C(T)/T$ remaining finite at absolute zero, which implied that the entropy itself defined by $ds = [ \: C(T)/T \: ] \: dT$ remained finite at $0$~K.

In response to Einstein's objections to the heat theorem, expressed during the First Solvay Congress in Brussels in the fall of 1911, \cite{Nernst_1912} suggested the alternative principle of unattainability of $T=0$ (``Unerreichbarkeit des absoluten Nullpunktes''), namely that absolute zero temperature cannot be reached in a finite time interval and in a finite number of steps.
A proof for this second way to express the third law of thermodynamics is given in the recent paper \citet{Masanes_Oppenheim_2017}.

Planck made the heat theorem more general in his third (1911) and fifth (1917) German editions of his Treatise on Thermodynamics \citep{Planck_1917}.
The third formulation of the third law expressed by Planck was clearly based on the entropy and was written in two parts:
\begin{enumerate}
\vspace*{-2mm}
\item (page 273) ``The gist of the (heat) theorem is contained in the statement that, as the temperature diminishes indefinitely the entropy of a chemical homogeneous body of finite density approaches indefinitely near to a definite value, which is independent of the pressure, the state of aggregation and of the special chemical modification'';
\vspace*{-2mm}
\item (page 274) ``without loss of generality, we may write (this definite value) $\lim \; (T\rightarrow \: 0) $ entropy $ \: = 0$.
We can now in this sense speak of an absolute value of the entropy''.
\end{enumerate}
\vspace*{-1mm}
The ``finite density'' hypothesis clearly excludes applications to perfect gases at $0$~K and the third law must only be applied to the solid state.

Planck wrote in the preface of the fifth edition in 1917: 
``The theorem in its extended form has in the interval (since 1911) received abundant confirmation and may now be regarded as well established''.
However, it was discovered after 1917 that a residual entropy may exist for anomalous species like Ice-Ih at $0$~K due to quantum effects explained by \citet{Pauling_1935} and \citet{Nagle_1966} (proton disorder and remaining randomness of hydrogen bonds at $0$~K).
For these reasons, Planck's formulation must now be amended and applied to the ``most stable form of pure crystalline solid substances''.

Accordingly, the standard third-law values of entropies at temperature $T_0$ and pressure $p_0$ can be written as
\vspace*{-1mm}
\begin{align}
\! \! \! \! \! \!
  s(T_0,p_0)  &  = \!
  \int^{T_0}_0 \frac{c_p(T,p_0)}{T} \; \; dT
   \: + \sum_k \: \frac{L(T_k,p_0)}{T_k}
\label{eq_s_third_law}
 \: .
\end{align}

\noindent The integral of the piecewise function $c_p(T,p_0)$ is computed from $0$~K to $T_0$ for all solid(s), liquid and gaseous states.
The sum over $k$ is extended to all transitions of phases occurring at $T_k$  with the latent heats $L(T_k)$.

The validity of the third law is now considered as established by the relevance of the computations of $\Delta G$, which provide accurate predictions of the constants of chemical reactions and of the spontaneity of these reactions with respect to temperature.
This is why thermochemical tables include absolute values for entropy $S$, whereas only relative standard enthalpies of reactions $\Delta H^0_r$ are provided.

According to the results already available in \citet{Kelley_1932} and in more recent Tables, the absolute values of entropies determined from the calorimetric method (\ref{eq_s_third_law}) and with the third law are in good agreement with the other available methods:
1) by using the statistical physics method, i.e., by computing the translational partition function and the corresponding Sackur-Tetrode equation valid for monoatomic gases only, and thus for Argon in the atmosphere \citep{Grimus_2013};
2) by computing the partition functions valid for more complex molecules (like O${}_2$, N${}_2$, H${}_2$O and CO${}_2$), including the impact of possible rotational, vibrational or anharmonic vibration corrections or electronic states of molecules \citep[and subsequent works]{Gordon_Barnes_1932,Gordon_34,Gordon_35};
3) by analyzing residual-ray frequencies and infrared absorption spectra of crystals or by using calculations from spectroscopic data for gases;
4) by measuring both $\Delta G$ and $\Delta H$, leading from Eq.~(\ref{eq_G_H_TS_Gibbs}) to 
$\Delta S = (\Delta H \: - \: \Delta G)/T$.

It is important to note that the third law corresponds to a similar hypothesis, which is implicitly assumed in Boltzmann's entropy written as $S = S_0 \: + \: k \: \ln(W)$, namely that $S_0$ is a constant, which is usually set to zero.
It is clearly explained in \citet[lectures delivered in 1944 at Dublin]{Schrodinger_1954} that the second part of the third-law assumption suggested by \citet{Planck_1917}, to write $S_0 = 0$ without loss of generality, must not be regarded as the essential thing.
This would create confusion and would draw attention away from the point really at issue, namely that $S_0$ is a universal constant that has no physical meaning only if it is independent of any internal physical parameters of any species, no matter what the value of $S_0$ might be.

The entropies of the five most frequent atmospheric species are listed in Table~A1 
for the values computed with either the third-law quantum statistical-physics (theoretical) method or with the third-law calorimetric (experimental) method.
Good agreement can be observed between theoretical and experimental values.
The larger discrepancy observed for H${}_2$O and GR96 ($45.132$ versus $44.31$) is not observed for MG15 ($45.2$).
The explanation for the larger value observed for O${}_2$ and MG15 ($49.7$ versus $49.031$) is given in Appendix~B.





%
%
%

\vspace{8mm}
\noindent
{\bf \underline{Appendix B: Applications of the third law} 
                 \underline{to atmospheric studies.}}
             \label{appendixB}
\renewcommand{\theequation}{B.\arabic{equation}}
  \renewcommand{\thefigure}{B.\arabic{figure}}
   \renewcommand{\thetable}{B.\arabic{table}}
      \setcounter{equation}{0}
        \setcounter{figure}{0}
         \setcounter{table}{0}
\vspace{2mm}

The apparent paradox is that, while there is no chemical reaction, the third law discovered in thermochemistry must be applied if the moist-air entropy is to be computed in the atmosphere.

The close link between the two subjects can be understood by analyzing the simple case of mixing dry air and water vapor, both of them considered as perfect gases.
If the gas constants $R_d$ and $R_v$ and specific heats $c_{pd}$ and $c_{pv}$ are assumed to be constant in the range of atmospheric temperatures (from $170$ to $340$~K), the entropies for dry air, water vapor and moist-air mixing can be written as
\vspace*{-1mm}
\begin{align}
\! \! \! 
    s_d(T,p_d)   &  =  c_{pd}  \ln(T/T_0) -  R_d \ln(p_d/p_{d0}) +  s_{d0}
  \label{eq_dry_air_entropy} \: , \\
\! \! \! 
    s_v(T,e)     & = \: c_{pv}   \ln(T/T_0) -  R_v  \ln(e/e_0)  + s_{v0}
  \label{eq_water_entropy} \: ,
\end{align}

\begin{align}
 \! \! \!   \! s(T,p,e,q_v) & = \: q_d \; s_d(T,p-e) \: + \: q_v \; s_v(T,e)
  \label{eq_moist_air_entropy1} \: , \\
\!  \! \! \!  s(T,p,e,q_v) & = \: c_{p} \ln(T/T_0) 
              \: - \: q_d \: R_d \: \ln[\:(p-e)/p_{d0}\:]
  \nonumber \\
       & \! \! \! \! \! \! \! \! \! \! - \: q_v \: R_v \: \ln(e/e_0)
         \: + 
   \left[ \: q_d \: s_{d0} \: + \: q_v \: s_{v0}  \: \right]
  \label{eq_moist_air_entropy2} ,
\end{align}
where $p=p_d+e$, $q_d=1-q_v$, $c_p = q_d \: c_{pd} \: + \: q_v \: c_{pv} $ and with the use of the partial pressures $p$, $e$, $p_{d0}$ and $e_0$, which automatically takes the entropy of mixing of the two gases into account.

The impact of the last bracketed terms in Eq.~(\ref{eq_moist_air_entropy2}) is similar to the impact of the last bracketed terms in Eq.~(\ref{eq_A_Nernst_1906_int}).
These terms are the product of the constants of integration $A(T_0)/T_0$, $s_{d0}$ and $s_{v0}$ by the variable terms $T(x,y,z)$, $q_d(x,y,z)$ and $q_v(x,y,z)$, respectively.

Therefore, the same consequences as described for thermochemistry hold true for the atmosphere:
in order to determine if a process is diabatic or isentropic, or for making isentropic analyses, it is necessary to compute local values of moist-air entropy with Eq.~(\ref{eq_moist_air_entropy2}) and, in particular, $q_d(x,y,z) \: s_{d0}$ and $q_v(x,y,z) \: s_{v0}$.
Consequently, the two reference values $s_{d0}$ and $s_{v0}$ must be known and the use of the third-law values would be in full agreement with general thermodynamics recommendations.


\citet{Bjerknes_1904} may well have been the first to imagine to using entropy in weather prediction, as a prognostic variable in one of his seven basic equations.
It is recalled in \citet{Marquet16a} that the consequences of the third law on atmospheric thermodynamics was described soon after in \citet[ p.158-160]{Richardson22}, who was already aware of the problem of the constant of integration in the entropy.
Richardson suggested that
``the most natural way of reckoning the entropy of the water substance would be to take it as zero at the absolute zero of temperature''.
This corresponds to the use of the third-law values for $s_{d0}$ and $s_{v0}$.

However, Richardson was not able to continue accurate computations of the moist-air entropy in 1922 because values of $c_p(T)$ were not available at that time for an absolute temperature varying from near-zero to $350$~K.
These measurements were made later on for all atmospheric species by using the magnetic refrigeration method to attain extremely low temperatures far below $1$~K \citep{Giauque49,Tiselius49}. 
It is now possible to find the absolute values of entropies for all atmospheric species: N${}_2$, O${}_2$, Ar, H${}_2$O, CO${}_2$, etc. in thermodynamic tables.
All these third-law values were already available in \citet{Kelley_1932}.


The use made of entropy by in \citet{Rossby_1937} for a moist-air isentropic analysis was based on the use of the dry-air value $\theta$. 
This seems unrealistic.

The first application of the third law for computing the moist-air entropies in atmospheric science was probably that of \citet{HaufHoller87}, who wrote that ``the entropy reference value 
($q_d \: s_{d0} \: + \: q_v \: s_{v0}$)
is not a constant'', and that ``the values of the zero-entropies have to be determined experimentally or by quantum statistical considerations'', namely from the third law or spectroscopic methods. 
They used the numerical values
$s_{v0} \approx 10320$~J~K${}^{-1}$~kg${}^{-1}$ and
$s_{d0} \approx 6675$~J~K${}^{-1}$~kg${}^{-1}$
for $T_0 = 273.15$~K and $p_{d0} = e_0 =1000$~hPa.

Another application of the third law to the atmosphere was that of \citet{Bannon_2005}, who used the reference entropies given by \citet{Chase_98} with the same value of $s_{v0}$ but with a smaller value $s_{d0} \approx 6612$~J~K${}^{-1}$~kg${}^{-1}$ for dry air.

The term $\Lambda_r$ appearing in Eq.~(\ref{eq_thetas}) depends on the standard values at $273.15$~K and $1000$~hPa considered in \citet{HaufHoller87} including a correction made in M11 to account for the impact of changes in partial pressures: $e_r \approx 6.11$~hPa for water vapor and $p_{dr} \approx 1000 - 6.11 \approx 994$~hPa for dry air.
The reference entropies are
$s_{vr} \approx 12673$~J~K${}^{-1}$~kg${}^{-1}$ and
$s_{dr} \approx 6777$~J~K${}^{-1}$~kg${}^{-1}$,
leading to $\Lambda_r = [ \: {(s_v)}_r - {(s_d)}_r \: ]/c_{pd} \approx 5.87$.

The third-law reference entropies given by \ref{eq_s_third_law} are explicitly computed in \citet{Marquet15a} and \citet{Marquet_Geleyn_2015} for N${}_2$, O${}_2$ and H${}_2$O.
A larger value $s_{d0} \approx 6848$~J~K${}^{-1}$~kg${}^{-1}$ for dry air is derived by considering the change in $c_p(T,p_0)$ for O${}_2$ at the second-order solid $\alpha$-$\beta$ transition occurring at $23.85$~K, forming a kind of Dirac pulse with no latent heat \citep{Fagerstroem_1969}.
The corresponding term $\Lambda_r \approx 5.80$ is $1.1$~\% smaller than the one considered in M11. 
The reference entropies considered in \citet{Chase_98} lead to a $2.7$~\% larger value, $\Lambda_r \approx 6.03$,
where the second order transition for O${}_2$ at $23.85$~K is not taken into account in JANAF Tables.

It is shown in \citet[section 2.4]{Marquet15a} and \citet[section 5.3]{Marquet_Geleyn_2015} that the linear combination $(1-a)\:s(\theta_e) + a\:s(\theta_l)$ described in Appendix~C of \citet{Pauluis_al_2010} can lead to the  third-law value of entropy $s(\theta_{s/M11})$ if the weighting factor is set to the value 
$a  = [ \: {(s_d)}_r - {(s_l)}_r \: ] \: T_r \: /L_v(T_r) \approx 0.356$.
Another value for $a$ would lead to a definition of the specific moist-air entropy which would disagree with the third law. 
The case $a=1$ corresponds to $S(\theta_l)$ if it is assumed that both ${(s_d)}_r$ and ${(s_v)}_r$ vanish at zero Celsius ($T=273.15$~K).
The case $a=0$ corresponds to $S(\theta_e)$ if it is assumed that both ${(s_d)}_r$ and ${(s_l)}_r$ vanish at zero Celsius.

The definitions of entropy of moist air selected in the IAPWS releases and guidelines (\url{http://www.iapws.org/}) are mostly based on reference-state conditions of vanishing entropies of both dry air and liquid water at the standard ocean state $273.15$~K and $1013.25$~hPa
\citep{IAPWS10,Feistel_2010}. 
However, an absolute (third-law) definition for entropy of ice-Ih is envisaged in \citet{Feistel_Wagner_2006}.
It is based on \citet{Gordon_34} and a value of $10320.7$~J~K${}^{-1}$~kg${}^{-1}$, which is close to that taken in \citet{HaufHoller87} and M11 for water vapor at $273.15$~K and $1000$~hPa.

The advantage of Eq.~(\ref{eq_s_thetas_M11}) is that both $s_{\rm ref}$ and $c_{pd}$ are constant, making $\theta_s$ a true equivalent of $s$.
Other definitions of $\theta_l$ or $\theta_e$, and even $\theta_S$ defined in \citet{HaufHoller87}, have the disadvantage of using values of $s_{\rm ref}$ and $c_{p}$ as factors of the logarithm, which both depend on $q_t$ or $r_t$, thus preventing these potential temperatures from being an equivalent of the moist-air entropy if $q_t$ is not a constant.


\bibliographystyle{ametsoc2014}
\bibliography{arXiv_Marquet_Isentropic_analyses_R2}


    \end{document}